\documentclass{article}
\usepackage[utf8]{inputenc}\DeclareUnicodeCharacter{2212}{-}
\usepackage{pgf}
\usepackage{import}

\usepackage[left=2.5cm,right=2.5cm]{geometry}

\let\lpol\l
\newcommand{\be}{\begin{equation}}
\newcommand{\ee}{\end{equation}}
\newcommand{\bea}{\begin{eqnarray}}
\newcommand{\eea}{\end{eqnarray}}

\newcommand{\Tr}{{\rm Tr}}

\renewcommand{\l}{\left}

\renewcommand{\Re}{\mathrm{Re} \,}

\def\eq#1{Eq.~(\ref{#1})}

\def\cyp{a}
\def\cyi{b}
\def\upi{c}
\def\hub{d}
\def\ToR{e}
\usepackage{array,multirow,hhline}
\usepackage{amsmath}
\usepackage{amssymb}
\usepackage{siunitx}
\usepackage{dsfont}
\usepackage{slashed}
\usepackage{mathtools}
\usepackage{braket}
\usepackage{float}

\usepackage{caption}
\usepackage[sorting=none,backend=bibtex,style=numeric-comp,natbib=true]{biblatex} 
\begin{document}

\begin{titlepage}
  \begin{center}
    \begin{LARGE}
      \textbf{Neutron electric dipole moment using lattice QCD simulations at the physical point } \\
    \end{LARGE}
  \end{center}

\vspace{.5cm}

\vspace{-0.8cm}
  \baselineskip 20pt plus 2pt minus 2pt
  \begin{center}
    \textbf{C.~Alexandrou\(^{(\cyp, \cyi)}\),
    A.~Athenodorou\(^{(\upi, \cyi)}\),
    K.~Hadjiyiannakou\(^{(\cyp, \cyi)}\)
    A.~Todaro\(^{(\cyp,\hub,\ToR)}\)
    }
  \end{center}
  
  \begin{center}
    \begin{footnotesize}
        \noindent
        \(^{(\cyp)}\) Department of Physics, University of Cyprus, P.O. Box 20537,
        1678 Nicosia, Cyprus\\
        \(^{(\cyi)}\) Computation-based Science and Technology Research Center, The Cyprus Institute, 20 Kavafi Str., Nicosia 2121, Cyprus \\
        \(^{(\upi)}\) Dipartimento di Fisica, Università di Pisa and INFN, Sezione di Pisa, Largo Pontecorvo 3, 56127 Pisa, Italy \\
        \(^{(\hub)}\) Institut für Physik, 
Humboldt-Universität zu Berlin, Newtonstr.\ 15, 
12489 Berlin, Germany\\
 \(^{(\ToR)}\)Dipartimento di Fisica, 
Università di Roma ``Tor Vergata'', 
Via della Ricerca Scientifica 1, 
00133 Rome, Italy\\
    \vspace*{-0.8cm}
    \end{footnotesize}
  \end{center}
  \centerline{\includegraphics[scale=0.19]{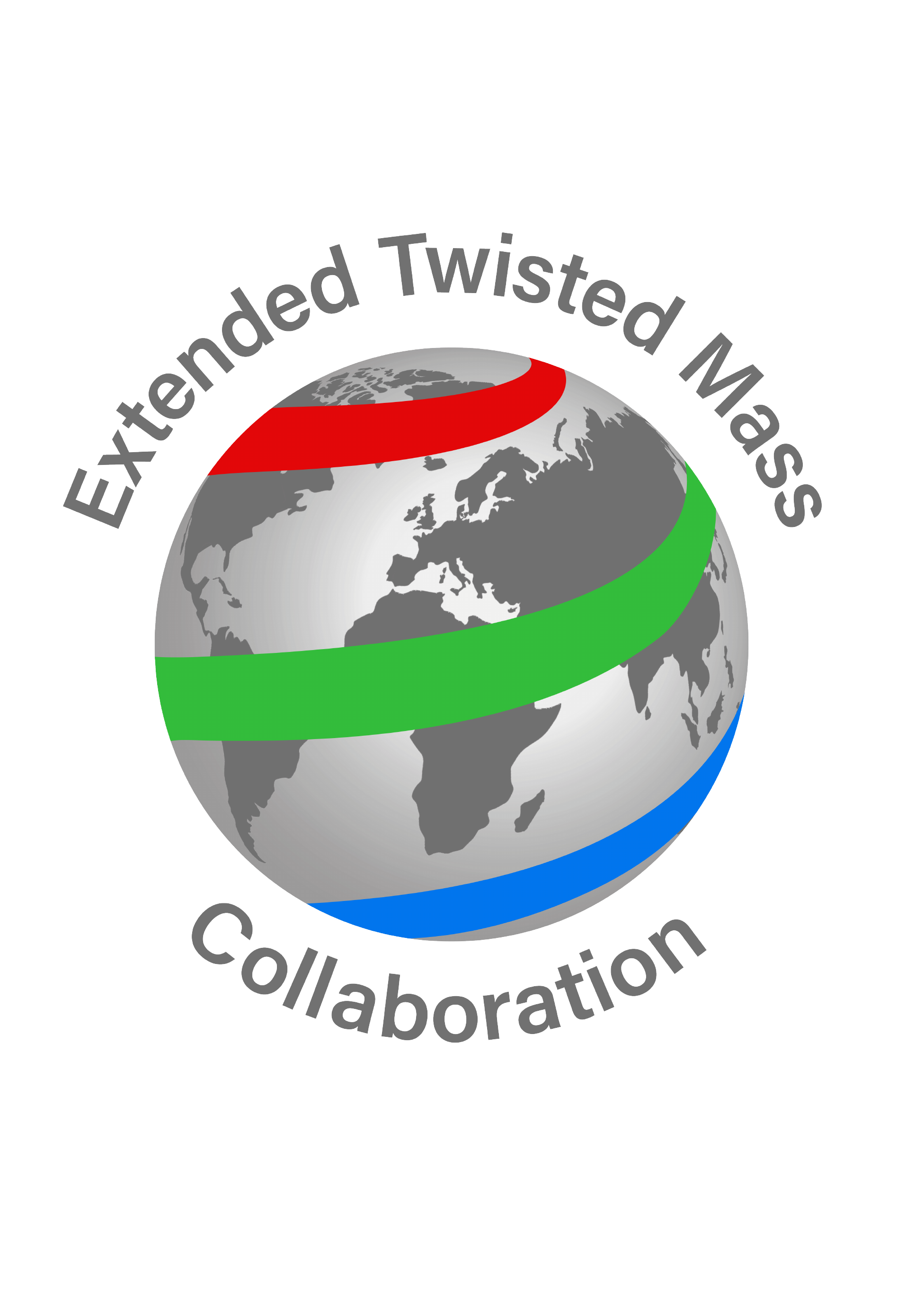}}
  \vspace*{-0.8cm}
  \begin{abstract} 
  We extract the neutron electric dipole moment  $\vert \vec{d}_N\vert$ within the lattice QCD formalism. We analyse one ensemble of $N_f=2+1+1$ twisted mass clover-improved fermions  with  lattice spacing of $a \simeq 0.08 \ {\rm fm}$ and physical values of the quark masses corresponding to a pion mass $m_{\pi} \simeq 139 \ {\rm MeV}$. The neutron electric dipole moment is extracted  by computing  the $CP$-odd electromagnetic form factor $F_3(Q^2 \to 0)$ through  small $\theta$-expansion of the action. This approach requires the calculation of the topological charge for which we employ a fermionic definition by means of spectral projectors while we also provide a comparison with the gluonic definition accompanied by the gradient flow. We show that using the topological charge from spectral projectors leads to absolute errors that  are more than two times smaller than those provided when the field theoretic definition is employed. We find a value of $\vert \vec{d}_N\vert =  0.0009(24) \ \theta \ e \cdot {\rm fm}$ when using the fermionic definition, which is statistically consistent with zero. 
  
  \begin{center}                                                                             \today
  \end{center}
  \end{abstract}
\end{titlepage}                                                                                                                     

\section{Introduction}
\label{sec:introduction}
The  discrete symmetries of parity $P$, charge conjugation $C$ and time-reversal $T$ play a significant role in the phenomenological structure of the Standard Model (SM). Although the neutron has an overall zero electric charge, a conundrum that physicists are trying to understand for decades is whether  the distribution of the positive electric charge could possibly not coincide with the distribution of the negative electric charge. That is to say whether there is no invariance under $CP$ parity, which can address the well known unsolved puzzle of the origin of the imbalance of matter and antimatter in the universe. Such an imbalance between positive and negative electric charge would manifest it self as the non-vanishing of the neutron  electric-dipole moment (nEDM). 

Up to the present, no finite nEDM value has been measured in experiments. In addition, reported current bounds are still several orders of magnitude below the SM prediction on $CP$ violation induced by the weak interactions. A finite  nEDM value would point towards physics   beyond the standard model (BSM)~\cite{Pospelov:2005pr} and it is thus an interesting quantity to study theoretically. 

Experimental measurements are under way to improve the upper bound of the value of nEDM denoted as $\vec{d}_N$ provided by a number of experiments, such as those given in Refs.~\cite{Helaine:2014ona,Baker:2006ts,Baker:2007df}. Until recently, the best measured upper bound was that 
given  in Refs.~\cite{Baker:2006ts,Baker:2007df} as
\be
\vert \vec{d}_N \vert  < 2.9 \times 10^{-13} e \cdot {\rm fm} \ (90\% \ {\rm CL})\,.
\label{eq:nEDM_smaller_experimental_upperbound}
\ee 
This result is extracted using  stored ``ultra-cold'' neutrons, applying a weak magnetic field when a
strong, parallel background electric field is reversed and measuring the alternation on the neutron spin precession frequency. The experiment has been carried out  at the Institut Laue-Langevin (ILL) reactor in Grenoble. 

Very recently~\cite{Abel:2020gbr},  an experiment measured the  nEDM at the Paul Scherrer Institute (PSI) in Switzerland using Ramsey's method of separated oscillating magnetic fields with "ultra-cold" neutrons. The novelty of this experiment lies on using the Hg-199 co-magnetometer and an array of optically pumped cesium vapor magnetometers to cancel and correct for magnetic field changes.  The result of this experiment yields  an improved upper limit of
\be
\vert \vec{d}_N \vert  < 1.8 \times 10^{-13} e \cdot {\rm fm} \ (90\% \ {\rm CL})\,.
\label{eq:nEDM_recent_value}
\ee 
The  nEDM value has also been extracted from different theoretical perspectives, such as model dependent studies \cite{Baluni:1978rf,Crewther:1979pi,Shifman:1979if,Schnitzer:1983pb,Shabalin:1982sg,Musakhanov:1984qy,Cea:1984qv,Morgan:1986yy,McGovern:1992bk},
as well as effective field theory calculations \cite{Pich:1991fq,Borasoy:2000pq,Hockings:2005cn,Narison:2008jp,Ottnad:2009jw,deVries:2010ah,Mereghetti:2010kp,deVries:2012ab,Guo:2012vf},
reporting values for  $d_N$ in the range of $\vert { \vec{d}_N} \vert \sim \theta \cdot {\cal O} \left( 10^{-2} - 10^{-3} \right) e \cdot {\rm fm}$. Using either the result given in \eq{eq:nEDM_smaller_experimental_upperbound} or \eq{eq:nEDM_recent_value} one derives a bound of the order $\theta \lesssim {\cal O} \left( 10^{-10} -  10^{-11} \right)$.

In this work, we investigate the nEDM induced by strong interactions, namely by the topological term. We neglect contributions which might come from higher dimensional components in the action, such as the lowest-dimensional ($d=5$) effective quark-gluon $CP$-odd interaction~\cite{Syritsyn:2019vvt,Syritsyn:2018mon}. In this approach, we include a $\theta$-topological term in the QCD Lagrangian density:
\bea
  {\cal L}_{\rm QCD} \left( x \right)= \frac{1}{2 g^2} 
{\rm Tr} \left[ F_{\mu \nu} \left( x \right) F_{\mu \nu} \left( x \right) \right] + 
\sum_{f} {\overline \psi}_{f} \left( x \right) (\gamma_{\mu} D_{\mu} + m_f) \psi_{f}\left( x \right)- i \theta q \left( x \right)\,,
\label{eq:QCD_Lagrangian_theta}
\eea
written in Euclidean time. The first two terms are $CP$-conserving while the $\theta$-term is $CP$-violating that can give rise to a non-zero nEDM. In the above expression, $\psi_f$ denotes a fermion field of flavor $f$ with bare mass $m_f$, $F_{\mu \nu}$ is the gluon field tensor and $q \left( x \right)$ is the topological charge
density, which in Euclidean space, is defined as 
\bea
  q \left( x \right) = \frac{1}{ 32 \pi^2} \epsilon_{\mu \nu \rho \sigma} 
{\rm Tr} \left[ F_{\mu \nu} \left( x \right) F_{\rho \sigma} \left( x \right) \right]\,.
\label{eq:Topological_Charge_Density}
\eea

By considering the electroweak (EW) sector of the SM, the action in \eq{eq:QCD_Lagrangian_theta} receives a contribution from the quark mass matrix $M$, arising from the Yukawa couplings to the Higgs field.
Hence, the parameter $\theta$ shifts to ${\overline \theta} = \theta + \arg\det M$ where now ${\overline \theta}$ describes the $CP$-violating parameter of the extended strong and EW symmetry. Here a delicate issue arises, namely that given the smallness of the total value of ${\overline \theta}$, either $\theta$ and $\arg\det M$ are both tiny or they cancel each other at the level that  satisfies the experimental bound on the nEDM value. This is referred to as the ``strong $CP$ problem''. We assume that a near cancellation of $\theta$ and $\arg\det M$ is extremely difficult to occur, thus, we take $\theta$ to be small and we perturbatively expand around it.

At low momentum transfer, the nucleon effective Lagrangian gives rise to the expression for the nEDM~\cite{Pospelov:2005pr}:
\bea
\vert \vec{d}_N \vert =  \theta \lim_{Q^2 \to 0} \frac{\vert F_3(Q^2) \vert}{2 m_N}\,,
\label{eq:dN}
\eea
at leading order in $\theta$, where  $m_N$ denotes the mass of the neutron, $Q^2{=}-q^2$ the four-momentum transfer in Euclidean space ($q{=}p_f-p_i$) and $F_3(Q^2)$ is the $CP$-odd neutron form factor.
We can, therefore, calculate the electric dipole moment by extracting the zero momentum transfer limit of the $CP$-odd form factor. This is the framework on which our work is based on. In practice, it is impossible to extract $F_3(Q^2)$ at $Q=0$ due to the fact that the $CP$-violating nucleon matrix element, decomposes to $Q_k F_3(Q^2)$ ($k{=}1,2,3$) and not just to $F_3(Q^2)$. Hence,  a direct extraction of $F_3(0)$ is prohibited by the theory. We then need to parametrize the $Q^2$-dependence of $F_3(Q^2)$ and then take the limit   $Q^2= 0$.

Since we expect that the $\theta$-parameter is very small we can expand the exponential of  the topological term in powers of $\theta$. This enables us to integrate out (over space-time) the topological charge density, which gives the total topological charge ${\cal Q}$:
\be
e^{i \theta \int d^4x \,q(x)} \equiv e^{i \theta {\cal Q}} = 1 + i \theta {\cal Q} + {\it O}(\theta^2)\,; \quad \quad {\cal Q}=\int  d^4x \,q(x) \,.
\label{eq:exp_eitheta}
\ee

A consequence of this expansion is that the value of $F_3(Q^2)$ depends on  the nucleon two- and three-point functions correlated with the  topological charge ${\cal Q}$, namely  with $\langle J_N ({\vec p}_f,t_f) 
{\overline J}_N({\vec p}_i,t_i) {\cal Q} \rangle$  as well as $\langle J_N ({\vec p}_f,t_f) {\cal J}^{\rm em}_\mu(\vec q, t) 
{\overline J}_N({\vec p}_i,t_i) {\cal Q} \rangle$, respectively. While the computation of nucleon two- and three-point functions within lattice QCD in the absence of the parity- violating term is well-known and rather precise~\cite{Alexandrou:2020sml,Alexandrou:2019brg,Alexandrou:2018sjm}, when correlating with ${\cal Q}$, introduces large statistical fluctuations. The topological charge, if sampled adequately, has a gaussian distribution centered at zero with a width that increases with the volume of the lattice and would by itself average to zero. If there is no correlation, or mild correlation, between the correlation functions and the topological charge then no signal can be obtained for $F_3(Q^2)$. Different definitions of the topological charge may be more suitable to pick up the correlations with the nucleon two- and three-point functions~\cite{Alexandrou:2017bzk,Alexandrou:2017hqw}. 

Recently, a comparison of lattice QCD determinations of the topological susceptibility using a gluonic definition with gradient-flow, cooling and other equivalent smoothing schemes, as well as, a fermionic definition using spectral projectors~\cite{Luscher:2004fu,Luscher:2010ik,Giusti:2008vb,Bonanno:2019xhg} has been carried out~\cite{Alexandrou:2017bzk,Alexandrou:2017hqw}. Although both these gluonic and fermionic definitions lead to compatible results in the continuum limit,  the gluonic definitions are much more affected by cut-off effects. 
For the spectral projectors one can choose a value of the spectral cutoff that would completely eliminate discretization effects in the topological susceptibility. Therefore, using spectral projectors is preferable since the discretization error    on the topological susceptibility as well as the statistical error are suppressed.  Given the statistical superiority of spectral projectors in calculation of quantities that depend on the topological charge, one needs to consider the fermionic definition as compared to the commonly used gluonic definition. One has to bear in mind that the computational cost of spectral projectors is orders of magnitude larger than that of the gluonic. However,  for quantities calculated at a given lattice spacing without a continuum extrapolation, the fermionic definition can be the only option to avoid  large cut-off effects. Since in this work we investigate the nEDM at physical pion mass we are, at the moment, restricted to use one $N_f=2+1+1$ ensemble~\cite{Alexandrou:2018egz}, and therefore a continuum extrapolation is not possible. In the future, we plan to include two more ensembles with smaller lattice spacings in order to be able to take the continuum limit.

An alternative way to compute the nEDM  is to use an expression from chiral perturbation theory to extrapolate the nEDM obtained at larger pion masses to the physical point.  However,  such  an extrapolation in the pion mass  carries its own uncontrolled systematic errors. Current values of $F_3(m_{\pi})$ computed for $m_{\pi}>  135$~ MeV within lattice QCD, as can be seen in Fig. ~\ref{fig:comparison},  are non-zero only within a few standard deviations or even consistent  with zero within the given statistical accuracy. Furthermore, $F_3$ decreases as the pion mass decreases vanishing at the chiral limit. Thus, the smaller the pion mass the more difficult it is to compute $F_3(Q^2)$  with a statistical significance that would exclude a zero value. Therefore, performing a chiral extrapolation from such data can be difficult. In a recent study~\cite{Dragos:2019oxn}, the authors performed a chiral fit and obtained   a statistically significant non-zero value for $F_3(0)$ at pion mass $m_{\pi}=135$~MeV, concluding that there is a signal for $CP$ violation induced by the $\theta$-term. However, large uncertainties on the few data points used in the fit, cast in doubt  the reliability of the result. It is, therefore, very important to perform a first-principles study directly at the physical pion mass.

This paper is organized as follows: In Section~\ref{sec:method} we explain our methodology, which enables us to extract $F_3(Q^2)$ from two and three-point correlation functions. Subsequently, in Section~\ref{sec:lattice_setup} we provide the details of the lattice QCD setup and the calculation, including the  expressions of the correlation functions in terms of the topological charge and discussion on the topological properties (Section ~\ref{sec:topCharge}) of the ensemble used in this work. In Section~\ref{sec:results} we present our results for the nucleon mixing angle and the $CP$-odd form factor $F_3(Q^2)$ as we take the $Q^2=0$ limit. We compare with other studies in Section~\ref{sec:comparison} and finally, in Section~\ref{sec:conclusions} we conclude.

\section{Method}
\label{sec:method}

The precise computation of the nEDM from first principles is one of the long-standing challenges and an active topic of research within  lattice QCD. The lattice QCD formulation provides an ideal framework  to access non-perturbatively the nEDM. The first pioneering attempt was reported nearly three decades ago~\cite{Aoki:1989rx}, and was based on the introduction of an external electric field and the measurement of the shift in the associated energy. This approach of extracting the nEDM within lattice QCD, however, breaks unitary since it requires the introduction of an external electric field. Subsequently,  two new approaches have been proposed. The most commonly used method involves the calculation of the $CP$-odd $F_3(Q^2)$ form factor by treating the $\theta$-parameter as a small perturbation~\cite{Guadagnoli:2002nm,Faccioli:2004jz,Shintani:2005xg,Shintani:2014zra,Shindler:2015aqa,Dragos:2019oxn}. Another approach is to extract the $CP$-odd $F_3(Q^2)$ form factor by simulating the theory with an imaginary $\theta$~\cite{Aoki:2008gv,Guo:2015tla}  to avoid the sign issue that a real $\theta$ introduces. This can be achieved either by using a field theoretical definition of the topological charge density or by replacing the topological charge operator with the flavour-singlet pseudoscalar density employing  the axial chiral Ward identities~\cite{Guadagnoli:2002nm,Faccioli:2004jz}. Although this approach provides a well-defined framework, it requires the production of new ensembles at various values of $\theta$ and analytic continuation. The cost of simulations of ensembles at the physical point for several values of $\theta$ and different values of lattice spacing is currently prohibitively high. Therefore, in this work we use the former approach, keeping $\theta$ real and expanding in powers of $\theta$.
\newpage

Allowing for the $CP$-violating $\theta$ term, the matrix element of the electromagnetic current 
\begin{equation}
    \braket{ N(p',s') | {\cal J}_{e.m.}^{\mu} | N(p,s)}_{\theta} = \bar{u}^{\theta}_N(p',s') \left[ \Gamma^\mu(q) \right] u^{\theta}_N(p,s)\,,
    \label{eq:mat_ele}
\end{equation}
can be decomposed in four form factors as follows
\begin{equation}
    {\Gamma}^\mu(q) = F_1(Q^2)\gamma^\mu + \left( F_2(Q^2) + i \gamma_5 F_3(Q^2)\right) \frac{i \sigma^{\mu\nu} q_{\nu}}{2 m_N^\theta}  + F_A(Q^2)\frac{(\slashed{q}q^\mu-q^2\gamma^\mu)\gamma_5}{m^{\theta,2}_N}\,.
    \label{eq:ff_def}
\end{equation}
 The electromagnetic current is given by \({\cal J}_{e.m.}^{\mu} = \sum_f e_f \bar{\psi}_f\gamma^{\mu}\psi_f \), where    \(e_f\)  is the electric charge of the quark field \(\psi_f\), \(\bar{u}^{\theta}_N(p',s')\) is the nucleon spinor in the presence of the $\theta$-term and \(q=p'-p\) is the four-momentum transfer. 
In the above expression \(F_1(Q^2)\) and \(F_2(Q^2)\) are the Dirac and Pauli electromagnetic form factors, \(F_3(Q^2)\) is the $CP$-odd form factor and \(F_A(Q^2)\) is the anapole form factor, that vanishes for $C$-preserving actions. They are all expressed as function of the Euclidean four-momentum transfer squared \(Q^2\). The presence of \(\braket{\quad}_{\theta}\) in Eq.~\eqref{eq:mat_ele} indicates that it is the state with the action that includes the \(\theta\)-term.
Once the dependence of the form factors on the transferred momentum is known, one can extract the electric dipole moment from \(F_3(Q^2)\) in the limit \(Q^2\rightarrow 0 \)  using the relation given in Eq.~\eqref{eq:dN}.

As pointed out in Ref.~\cite{Abramczyk:2017oxr}, however, a spurious mixing between the \(F_2(Q^2)\) and \(F_3(Q^2)\) form factors occurs if these quantities are not carefully defined. In particular, the contribution to the \({\cal J}^\mu_{e.m.}\) matrix elements coming from \(F_3(Q^2)\) transforms as an axial 4-vector as expected, only if the spinor \(u^{\theta}_N(p,s)\) appearing in Eq.~\eqref{eq:mat_ele} transforms as a regular Dirac spinor under parity, i.e. only if \(u^{\theta}_N(p'=(p_0,-\vec{p})) = \gamma_4 u^{\theta}_N(p)\). This holds if spinors satisfy the Dirac equations:
\begin{equation}
    \left(i\slashed{p} + m_{N}^{\theta} \right) u_N^{\theta}(p,s) = 0\,, \quad \quad \bar{u}_N^{\theta}(p,s) \left(i\slashed{p} + m_{N}^{\theta} \right) = 0\,,
    \label{eq:Dirac_eq}
\end{equation}
where the real-valued \(m_N^\theta\) is the mass of the nucleon in the \(\theta\neq 0\) vacuum.

On the lattice, the above matrix elements can be extracted from the Euclidean three-point function given by
\begin{equation}
 G_{3pt}^{\mu,(\theta)}(\vec{p}_f,\vec{q},t_f,t_{ins}) \equiv \braket{ J_N(\vec{p}_f,t_f) | {\cal J}_{e.m.}^{\mu}(\vec{q},t_{ins}) | \bar{J}_N(\vec{p}_i,t_i)}_{\theta} \,,
 \label{eq:G3pt_theta}
\end{equation}
where \(J_N(\vec{p}_f,t_f)\), \(\bar{J}_N(\vec{p}_i,t_i)\) are the nucleon interpolating operators that respectively create a nucleon at time \(t_i\) (source) with momentum \(\vec{p}_i\) and annihilate it at time \(t_f\) (sink) and momentum \(\vec{p}_f\). 

Inserting unity as a complete set of energy and momentum eigenstates in Eq.\eqref{eq:G3pt_theta}  leads, in the large Euclidean time limit, to the ground state contribution given as 
\begin{align}\nonumber
    G_{3pt}^{\mu,(\theta)}(\vec{p}_f,\vec{q},t_f,t_{ins},t_i) =\; & e^{-E_N^f(t_f-t_{ins})}e^{-E_N^i(t_{ins}-t_{i})} \\ 
    &\sum_{s,s'} \braket{J_N|N(p_f,s')}_{\theta} \braket{N(p_f,s')| {\cal J}_{e.m.}^{\mu}|N(p_i,s)}_{\theta} \braket{N(p_i,s)|\bar{J}_N}_{\theta}\,,
    \label{eq:gs_exp1}
\end{align}
with \(E_N^i\equiv E_N(\vec{p}_i) = \sqrt{\vec{p}_i^2 + (m_N^{\theta})^2 }\) and \(E_N^f\equiv E_N(\vec{p}_f) = \sqrt{\vec{p}_f^2 + (m_N^{\theta})^2 }\). Excited states contribution are  suppressed in the above expressions. The overlap between the interpolating operators and the nucleon state of a given momentum can be expressed as

\begin{equation}
\braket{J_{N}|N(\vec{p},s)}_{\theta} = Z_{N}^{\theta} \tilde{u}_N^{\theta} (\vec{p},s) \,, \quad \quad 
\braket{N(\vec{p},s)|\bar{J}_{N}}_{\theta} = (Z_{N}^{\theta})^* \bar{\tilde{u}}_N^{\theta} (\vec{p},s)\,,
\label{eq:eq_overlap}
\end{equation}
where the spinors \(\tilde{u}^{\theta}_{N}\),\(\bar{\tilde{u}}^{\theta}_{N}\) satisfy the following Dirac equations 
\begin{equation}
    \left(i\slashed{p} + m_{N}^{\theta} e^{-2i\alpha_N(\theta)\gamma_5}\right) \tilde{u}_N^{\theta}(\vec{p},s) = 0\,, \quad\quad \bar{\tilde{u}}_N^{\theta}(\vec{p},s) \left(i\slashed{p} + m_{N}^{\theta} e^{-2i\alpha_N(\theta)\gamma_5}\right) = 0\,.
    \label{eq:Dirac_eq_alfa}
\end{equation}
The imaginary phase in the mass terms that arises due to the $CP$-violation induced by the \(\theta\)-term, is parametrized by the so called mixing angle \(\alpha_N(\theta)\). These spinors are related to the \(u_N^\theta\), \(\bar{u}_N^\theta\) that are well-behaved under parity by an axial rotation, namely
\begin{equation}
    \tilde{u}_N^{\theta} = e^{i \alpha_N \gamma_5} u_{N}^{\theta}\,, \quad \quad    \bar{\tilde{u}}_N^{\theta} =  \bar{u}_{N}^{\theta} e^{i \alpha_N \gamma_5} \,.    
    \label{eq:ax_rot_tuu}
\end{equation}
This can be verified by substituting Eq.~\eqref{eq:ax_rot_tuu}~in Eq.~\eqref{eq:Dirac_eq_alfa} and noticing that one recovers Eq.~\eqref{eq:Dirac_eq}. With this in mind, one can rewrite Eq.~\eqref{eq:gs_exp1} as

\begin{align}\nonumber
    G_{3pt}^{\mu,(\theta)}(\vec{p}_f,\vec{q},t_f,t_{ins},t_i) \simeq\; & |Z_{N}^{\theta}|^2 e^{-E_N^f(t_f-t_{ins})}e^{-E_N^i(t_{ins}-t_{i})} \\ 
    & e^{i \alpha_N \gamma_5} \left(\frac{-i \slashed{p}_f + m_N^\theta}{2 E_N^f}\right) \Gamma^\mu(q) \left(\frac{-i \slashed{p}_i + m_N^\theta}{2 E_N^i}\right) e^{i \alpha_N \gamma_5}\,,
    \label{eq:gs_exp2}
\end{align}
 where we have used the summation property of the spinors.

For small \(\theta\), the quantities in the r.h.s of Eq.\eqref{eq:gs_exp2} can be safely replaced by their leading-order terms, as follows:
\begin{gather} \nonumber
m_N^{\theta} \simeq m_N + \mathcal{O}(\theta^2)\,, \quad \quad Z_N^{\theta} \simeq Z_N + \mathcal{O}(\theta^2) \, ,\\ \alpha_N^{\theta} \simeq \alpha_N^{(1)} \theta + \mathcal{O}(\theta^3) \quad,\quad F_3(Q^2) \simeq F_3^{(1)}(Q^2)\theta + \mathcal{O}(\theta^3),
\label{eq:theta_expansions}
\end{gather}
while higher order contributions can be neglected, therefore in the following sections we will simply refer to the mixing angle and the $CP$-odd form factor as $\alpha_N$ and $F_3(Q^2)$ correspondingly. The expectation values \(\braket{\dots}_\theta\) can also be expanded by using Eq.\eqref{eq:exp_eitheta}. This gives
\begin{equation}
\braket{\mathcal{O}}_{\theta} = \frac{1}{Z_{\theta}} \int [\text{dU}] [\text{d}\bar{\psi}] [\text{d}\psi] \: \mathcal{O} e^{-S_{QCD} +i \theta \mathcal{Q} } \simeq \braket{\mathcal{O}}_{\theta=0} + i\theta \braket{\mathcal{O}\mathcal{Q}}_{\theta=0} + \mathcal{O}(\theta^2).
\label{eq:pathInt_th}
\end{equation}
The left hand side (l.h.s.) of Eq.\eqref{eq:gs_exp2} is now written in terms of expectation values of states with \(\theta=0\) and  can be computed using gauge configurations extracted with the standard $CP$-even QCD action. The three-point function reads:
\begin{equation}
G_{3pt}^{\mu,(\theta)}(\vec{p}_f,\vec{q},t_f,t_{ins})    = G_{3pt}^{\mu,(0)}(\vec{p}_f,\vec{q},t_f,t_{ins}) + i \theta G_{3pt,\mathcal{Q}}^{\mu,(0)}(\vec{p}_f,\vec{q},t_f,t_{ins})\, + \cdots,
\label{eq:G3pt_0}
\end{equation}
with 
\begin{align}\label{eq:G3pt}
G_{3pt}^{\mu,(0)}(\vec{p}_f,\vec{q},t_f,t_{ins}) &\equiv \braket{ J_N(\vec{p}_f,t_f)  {\cal J}_{e.m.}^{\mu}(\vec{q},t_{ins})  \bar{J}_N(\vec{p}_i,t_i)}_{0} \,,\\  G_{3pt,\mathcal{Q}}^{\mu,(0)}(\vec{p}_f,\vec{q},t_f,t_{ins})  &\equiv \braket{ J_N(\vec{p}_f,t_f)  {\cal J}_{e.m.}^{\mu}(\vec{q},t_{ins})\, \mathcal{Q} \, \bar{J}_N(\vec{p}_i,t_i)}_{0}\,.
\label{eq:G3ptQ}
\end{align}
 One can, thus, divide the right hand side (r.h.s.) of Eq.~\eqref{eq:gs_exp2} into a \(CP\)-even and a \(CP\)-odd part and relate them respectively to the three-point functions of Eq.~\eqref{eq:G3pt} and Eq.~\eqref{eq:G3ptQ}. The unknown normalization coefficients \(Z_N\) can be canceled by taking an appropriate ratio with the two-point function, while the nucleon mixing angle \(\alpha_N\) can be measured by using the relation
\be
G_{2pt}^{(\theta)}(\vec{p}_f,t_f) \equiv \braket{ J_N(\vec{p}_f,t_f) \bar{J}_N(\vec{p}_i,t_i) }_{\theta} = \lvert Z^{\theta}_N \rvert^2 e^{-E_N(t_f-t_i)}\frac{-i\slashed{p}_f+m_N^{\theta}e^{i2\alpha_N^{\theta}\gamma_5}}{2 E_N}\,,
\label{eq:G2pt_theta}
\ee
and expanding  in powers of \(\theta\). The result is that the mixing angle \(\alpha_N\) can be determined from the computation of the following two-point functions
\begin{align}\label{eq:G2pt}
G_{2pt}^{(0)}(\vec{p}_f,t_f) &\equiv \braket{ J_N(\vec{p}_f,t_f) \bar{J}_N(\vec{p}_i,t_i)}_{0} \,,\\  G_{2pt,\mathcal{Q}}^{(0)}(\vec{p}_f,t_f)  &\equiv \braket{ J_N(\vec{p}_f,t_f)\, \mathcal{Q} \, \bar{J}_N(\vec{p}_i,t_i)}_{0}\,.
\label{eq:G2ptQ}
\end{align}
    
\section{Lattice setup}
\label{sec:lattice_setup}
We use one gauge ensemble of twisted mass fermions produced  with \(2\) degenerate light flavours, a strange and a charm quark ($N_F=2+1+1$), with all quark masses tuned close to their physical values. We use the Iwasaki improved gauge action~\cite{Iwasaki:1985we}, given by

\be
S_G = \frac{\beta}{3} \sum_x \left(c_0\sum_{ \substack{\mu,\nu=1 \\ \mu<\nu} }^4\left[ 1- \Re\Tr\left( U_{x,\mu\nu}^{1\times1}\right)\right] + c_1 \sum_{ \substack{\mu,\nu=1 \\ \mu\neq\nu} }^4 \left[ 1- \Re\Tr\left(U_{x,\mu\nu}^{1\times2}\right)\right]\right)\,,
\label{eq:g_action}
\ee
where \(\beta=6/g^2\), \(U^{1\times1}\) the plaquette and \(U^{1\times2}\) the rectangular Wilson loops. The Symanzik coefficients are set to \(c_0=3.648\) and \(c_1= (1-c_0)/8\).
The fermionic sector is implemented using the twisted mass formulation of lattice QCD \cite{Frezzotti:2000nk,Frezzotti:2003ni}, which for the degenerate light quark double  up and down has the form

\be
S_F^{l} = a^4\sum_x \bar{\chi}^{(l)}(x)\left( D_W[U] + \frac{i}{4} c_{SW}\sigma^{\mu\nu}\mathcal{F}^{\mu\nu}[U] + m_{0,l} + i \mu_l\gamma_5\tau^3 \right)\chi^{(l)}(x)\,.
\label{eq:fl_action}
\ee
In the equation above \(\chi^{(l)}\) is the field representing the light quarks doublet, expressed in the twisted basis, \(m_{0,l}\) and \(\mu_l\) are respectively the untwisted and twisted  mass parameters, \(\tau^3\) is the third Pauli matrix acting in flavor space and \(D_W\) is the massless Wilson-Dirac operator. The clover term \(\propto \sigma^{\mu\nu}\mathcal{F}^{\mu\nu}\) is included in the action to suppress cut-off effects reducing the difference between the mass of the charged and neutral pions~\cite{Alexandrou:2018egz}.
The strange and charm quarks are included as a non-degenerate twisted doublet \(\chi^{(h)}=(s,c)^t\), with the  action~\cite{Frezzotti:2003xj}
\be
S_F^{h} = a^4\sum_x \bar{\chi}^{(h)}(x)\left( D_W[U] + \frac{i}{4} c_{SW}\sigma^{\mu\nu}\mathcal{F}^{\mu\nu}[U] + m_{0,h} - \mu_{\delta}\tau^1 + i \mu_{\sigma}\gamma_5\tau^3 \right)\chi^{(h)}(x)\,,
\label{eq:fh_action}
\ee
 where \(m_{0,h}\) is the bare untwisted quark mass for the heavy doublet, \(\mu_{\delta}\) the bare twisted mass along the \(\tau^1\) direction and \(\mu_{\sigma}\) the mass splitting in the \(\tau^3\) direction.

We tune the partial conserved axial current (PCAC) mass to zero in order to achieve maximal twist. This ensures automatic \(\mathcal{O}(a)\) improvement for the expectation values of the observables of interest~\cite{Frezzotti:2005gi}. More details about the generation of this ensemble can be found in Ref.~\cite{Alexandrou:2018egz}. The lattice size is \(64^3\times128\), with lattice spacing \(a=0.0801(4) \text{fm}\), as determined from the nucleon mass \cite{Alexandrou:2018egz}, pion mass \(m_{\pi}=139(1) \text{MeV}\) and \(Lm_{\pi} = 3.62\). We will refer to this ensemble as  cB211.72.64. For the analysis, we used 750 gauge configurations, separated by 4 trajectories each. The parameters of the simulation are summarized  in Table~\ref{tab:params_sim}.

\begin{table}[H]
    \centering
    \begin{tabular}{c|c}
        \hline
        \hline
        \multicolumn{2}{c}{
        \(\beta=1.778,\qquad c_{SW}=1.69,\qquad a=0.0801(4)\)}\\
        \hline
         \(64^3\times 128\),  &  \(m_\pi=139(1) \text{MeV},\qquad m_\pi L=3.62\)\\
         \(L=5.13 \text{fm}\) &  \(m_N=940(2) \text{MeV}\)\\  
        \hline
        \hline
    \end{tabular}
    \captionof{table}{ Simulation parameters for the cB211.72.64 ensemble~\cite{Alexandrou:2018egz,Alexandrou:2018sjm} used in this work.}
    \label{tab:params_sim}
\end{table}

\subsection{Correlation functions}
\label{sec:correlation_functions}
For the computation of the nucleon two- and three-point functions we employ the standard proton interpolating field, namely 
\begin{equation}
    J_N(x) = \epsilon^{abc} \left[ u^{a,T}(x) \mathcal{C}\gamma_5 d^{b}(x) \right]\, u^{c}(x) \,,
    \label{eq:interp_fields}
\end{equation}
where \(u(x)\) and \(d(x)\) are up and down quark fields in the physical base, and \(\mathcal{C}=i\gamma_2\gamma_4\) is the charge conjugation matrix. Since up and down quarks are degenerate in our formulation, the proton and neutron are degenerate.
In order to improve the overlap with the neutron ground state and suppress excited states, we use Gaussian smeared quark fields~\cite{Gusken:1989qx,Alexandrou:1992ti}, with \(125\) smearing steps and parameter \(\alpha_G=0.2\). The smearing is tuned in order to approximately reproduce a mean square radius for the neutron  of 0.5~fm.  We apply \(50\) steps of APE-smearing \cite{Albanese:1987ds} on the gauge links that enter the smearing operator with \(\alpha_{\rm APE}=0.5\) in order to reduce gauge fluctuations.

For the electromagnetic current \({\cal J}^{\mu}_{e.m.}(x)\)  we use the symmetrized lattice conserved vector current defined as
\begin{align}\nonumber
j^{\mu}_f(x) = \frac{1}{4}&\left[\ \bar{\psi}_f(x+\hat{\mu}) U_{\mu}^{\dagger}(x)(1+\gamma_\mu) \psi_f(x) 
-\bar{\psi}_f(x) U_{\mu}(x)(1-\gamma_\mu) \psi_f(x+\hat{\mu})\right.\\
+&\left.\bar{\psi}_f(x) U_{\mu}^{\dagger}(x-\hat{\mu})(1+\gamma_\mu) \psi_f(x-\hat{\mu})
-\bar{\psi}_f(x-\hat{\mu}) U_{\mu}(x-\hat{\mu})(1-\gamma_\mu) \psi_f(x)\ \right] \,,
    \label{eq:J_em_noether}
\end{align}
which, in contrast to the local current \(\bar{\psi}_f(x)\gamma^{\mu}\psi(x)\), does not need renormalization.

\begin{figure}[h]
\begin{minipage}[b]{0.48\linewidth}
    \includegraphics[width=0.95\textwidth]{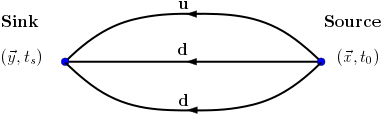}
\end{minipage}%
\hspace*{\fill}
\begin{minipage}[b]{0.48\linewidth}
    \includegraphics[width=0.95\textwidth]{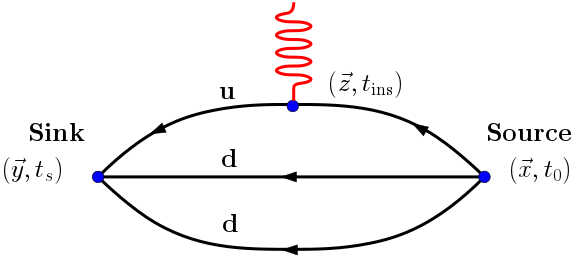}
\end{minipage}
\captionof{figure}{ Diagrammatic representation of the two-point function (left) and connected three-point function (right).}
\label{fig:diag_picture}
\end{figure}
The two- and three-point functions are then given by 
\begin{align}
    \label{eq:G2pt_prjcted}
    &G_{2pt}(\Gamma_0,\vec{p}_f,t_f,t_i) \equiv \sum_{\vec{y}} \text{Tr} \left[ \Gamma_0 \braket{ J_N(\vec{y},t_f) \bar{J}_N(\vec{x},t_i)} \right]  e^{-\vec{p}_f(\vec{y}-\vec{x})}\,,  \\
    \label{eq:G2ptQ_prjcted}
    &G_{2pt,\mathcal{Q}}(\gamma_5,\vec{p}_f,t_f,t_i) \equiv \sum_{\vec{y}} \text{Tr} \left[ \frac{\gamma_5}{4} \braket{ J_N(\vec{y},t_f)\, \mathcal{Q} \, \bar{J}_N(\vec{x},t_i)} \right]  e^{-\vec{p}_f(\vec{y}-\vec{x})}\,, \\ 
    \label{eq:G3pt_prjcted}
    &G^{\mu}_{3pt}(\Gamma_k,\vec{q},\vec{p}_f,t_f,t_{ins},t_i) \equiv \sum_{\vec{y},\vec{z}}  \text{Tr} \left[ \Gamma_k \braket{ J_N(\vec{y},t_f) \, {\cal J}_{e.m.}^{\mu}(z,t_{ins})\, \bar{J}_N(\vec{x},t_i)} \right] e^{-\vec{p}_f(\vec{y}-\vec{x})}e^{\vec{q}(\vec{z}-\vec{x})}\,,  \\ 
    \label{eq:G3ptQ_prjcted}
    &G^{\mu}_{3pt,\mathcal{Q}}(\Gamma_k,\vec{q},\vec{p}_f,t_f,t_{ins},t_i) \equiv \sum_{\vec{y},\vec{z}}  \text{Tr} \left[ \Gamma_k \braket{ J_N(\vec{y},t_f)\, {\cal J}_{e.m.}^{\mu}(z,t_{ins}) \, \mathcal{Q} \, \bar{J}_N(\vec{x},t_i)} \right] e^{-\vec{p}_f(\vec{y}-\vec{x})} e^{\vec{q}(\vec{z}-\vec{x})}\,,
\end{align}
where the projectors $\Gamma_0$ and $\Gamma_k$ are given by 
\begin{equation}
    \Gamma_0 = \frac{1}{4}(\mathds{1}+\gamma_0),\qquad \Gamma_k =  i\Gamma_0 \gamma_5\gamma_k\,.   
    \label{eq:projs}
\end{equation}

For the computation of the three-point functions, we consider  only the connected contribution as shown in Fig.~\ref{fig:diag_picture}. We use the standard method of  sequential inversions through the sink, taking the final momentum \(\vec{p}_f=\vec{0}\). The time slice of the sink relative to the source  (sink-source time separation) is kept fixed to \(t_{f}-t_{i}=12a\). From our previous investigation using an ensemble of $N_f=2+1+1$ twisted mass fermions with pion mass \(m_{\pi}\approx370\)~MeV we showed that such a time separation is sufficient to suppress excited state contributions to the accuracy of the present study~\cite{Alexandrou:2015spa}. Larger values of the sink-source time separation, even if desirable for a better suppression of the excited states, lead to large statistical uncertainties, that for the current investigation would require a prohibitively high statistics. 
We compute the three-point functions using \(54\) randomly distributed source positions per configuration, that can be treated as statistically independent measures, leading to a total statistics of \(\sim40\)k data.
The two-point functions entering the \(F_3\) determination both explicitly, appearing in the ratio with the three-point function (see Eq.~\ref{eq:Pi_munu} in Section~\ref{sec:cp_odd_form_factor}), and implicitly through the mixing angle \(\alpha_N\). In the former case, we employ the same \(54\) source positions used for the three-point functions, while for the computation of the mixing angle, we had higher statistics available, namely \(200\) source positions per configuration. A summary of the statistics used is given in Table~\ref{tab:stat}.

\begin{table}[h]
    \centering
    \begin{tabular}{c|c|c|c|c}
    \hline\hline
      & Correlation functions & \(N_{\rm src}\) & \(N_{\rm cnfs}\) & \(N_{\rm tot}\) \\
     \hline
         \(\alpha_N\) &  \(G_{2pt}\) & 200 & 750 & 150000\\
         \(F_3\) &  \(G_{2pt}\),\(G_{3pt}\) & 54 & 750 & 40500\\
    \hline\hline
    \end{tabular}
    \captionof{table}{Statistics employed for the evaluation of \(\alpha_N\) and \(F_3\).}
    \label{tab:stat}
\end{table}

\section{Topological charge}\label{sec:topCharge}

As already mentioned, for the computation of the three- and two- point functions in Eqs.~\eqref{eq:G2ptQ_prjcted},\eqref{eq:G3ptQ_prjcted} one needs the topological charge.  In the continuum, the topological charge is defined as the integral over the four-dimensional volume of the topological charge density given in Eq.~\eqref{eq:Topological_Charge_Density}, namely
\be
{\cal Q} = \frac{1}{32\pi^2} \int d^4 x \: \epsilon_{\mu\nu\rho\sigma} \Tr\left[F_{\mu\nu}(x)F_{\rho\sigma}(x)\right] \,.
\label{eq:Q_continuum_def}
\ee
The discrete counterpart of the above quantity can be obtained by replacing the gluonic field tensor with a lattice operator that reproduces the correct continuum limit. The choice is not unique, and operators with better finite-size effects can be obtained by using \({\cal O}(a)\)-improved discretizations of \(F_{\mu\nu}\). In particular, one of the definitions of \({\cal Q}\) we used in this work, is the symmetric or 'clover' definition, firstly introduced in Ref.~\cite{DiVecchia:1981aev}. It has the form
\begin{equation}
    \mathcal{Q}_L = \frac{1}{32\pi^2} \sum_{x} \epsilon_{\mu\nu\rho\sigma} \text{Tr}\left[C_{\mu\nu}(x)C_{\rho\sigma}(x)\right]\,,
    \label{eq:Q_fieldteo_def}
\end{equation}
and uses a discretization of the gauge strength tensor in terms of a "clover leaf" path \(C_{\mu\nu}\), made by the sum of the plaquettes \(P_{\mu\nu}(x)\) centered in \(x\) and with all the possible orientations in the \(\mu\nu\)-plane, i.e.
\begin{equation}
    C_{\mu\nu}(x) = \frac{1}{4}\text{Im}\left[P_{\mu\nu}(x)+P_{\nu,-\mu}(x)+P_{-\mu,-\nu}(x)+P_{-\nu,\mu}(x)\right]\,.
    \label{eq:clover}
\end{equation}
This operator is even under parity transformations and exhibits \(\mathcal{O}(a^2)\) discretization effects. We use the gradient flow \cite{Luscher:2010iy} in order to suppress the UV fluctuations of the gauge field defining the topological charge. 
The smoothing action employed in the flow equation is the standard Wilson action. The elementary integration step is \(\epsilon=0.01\) and the topological charge is computed on the smoothed fields at multiples of \(\Delta \tau_{\rm flow}=0.1\).
The flow time must be chosen large enough such that discretization effects are canceled but  to keep the topological properties of the gauge field unchanged. For this reason we study the dependence of our final quantities on \(\tau_{\rm flow}\) searching for a plateau region. Accordingly to Ref.~\cite{Luscher:2010iy} we expect that this happens for \(a\sqrt{8\tau_{\rm flow}}\sim \mathcal{O}(0.1\text{fm})\).

In addition, we use a second definition of the topological charge  based on spectral projectors as described in Refs.~\cite{Giusti:2008vb,Luscher:2010ik}. This  definition allows one to extract the topological charge from the spectrum of the hermitian Wilson-Dirac operator \(D_W^\dagger D_W\), by employing the relation
\begin{equation}
    \mathcal{Q}_0 = \sum_{i}^{ \lambda_i < M_0^2 } u_i^{\dagger} \gamma_5 u_i\,,
    \label{eq:fQ_bare}
\end{equation}
where \(u_i\) is the eigenvector related to the \(i\)-th eigenvalue \(\lambda_i\), \(\mathcal{Q}_0\) is the bare topological charge, and \(M_0\) is the bare spectral threshold. It bounds the modes that enter into the sum in Eq. \eqref{eq:fQ_bare} by requiring \(\lambda_i<M_0^2\). Accordingly to Ref.~\cite{Giusti:2008vb}, the renormalization of these quantities are given by
\begin{equation}
    \mathcal{Q} = \frac{Z_S}{Z_P}\mathcal{Q}_0 \:,\qquad M_{\rm thr} = Z_P^{-1} M_0\,,   
    \label{eq:fQ_ren}
\end{equation}
where \(Z_P\), \(Z_S\) are the renormalization constants of the pseudoscalar and scalar densities, respectively.  
We calculate the lowest 200 eigenvalues of the squared twisted mass Dirac operator using the Implicitly Restarted Lanczos Method (IRLM) where polynomial acceleration is employed. 

The renormalization constants are computed in a massless renormalization scheme, employing the Rome-Southampton method or the so-called RI$^\prime$ scheme~\cite{Martinelli:1994ty}. We use five  $N_f=4$ ensembles generated exclusively for the renormalization program at the same $\beta$ value as that of the  ensemble of interest.
The five ensembles are generated at different pion masses in the range of 366~MeV to 519~MeV using a lattice volume of $24^3 \times 48$. This enables us  to perform the chiral extrapolation to extract  the $Z$-factors at the chiral limit~\cite{Alexandrou:2019brg}.
  For the non-perturbative calculation of the vertex functions we use momentum sources~\cite{Gockeler:1998ye} that allow us to reach per mil statistical accuracy with ${\cal O}(10)$ configurations~\cite{Alexandrou:2010me,Alexandrou:2012mt}. 
For the renormalization we need $Z_P$ and $Z_S$, which are scheme and scale dependent. We use the ${\overline{\rm MS}}$-scheme, which is commonly used in experimental and phenomenological studies. The conversion procedure is applied on the $Z$-factors obtained on each initial RI$'$ scale $(a\,\mu_0)$, with a simultaneous evolution to an $\overline{\rm MS}$ scale, chosen to be $\overline{\mu}{=}$2 GeV. For the conversion and evolution we employ the intermediate Renormalization Group Invariant (RGI) scheme, which is scale independent and connects the $Z$-factors between the two schemes. For more details about  our renormalization program see Refs~\cite{Alexandrou:2019ali,Alexandrou:2019brg,Alexandrou:2020sml}. We find that $Z_P=0.462(4)$~\cite{Alexandrou:2019brg} and $Z_S=0.620(4)$~\footnote{We would like to thank M. Constantinou for providing this value.}.
 
Using these values of the renormalization constants, and changing the upper limit in Eq.~\eqref{eq:fQ_bare} up to the maximum 200 lowest eigenvalues corresponds to a threshold \(M_{\rm thr}\) that varies in the range \(0\div65\ \text{MeV}\). 

In the rest of the paper, we will refer to the definition of Eq.~\eqref{eq:Q_fieldteo_def} as ``gluonic" or ``field theoretic" definition of the topological charge, while the one defined by Eq.~\eqref{eq:fQ_ren} will be referred to as  the ``fermionic" or ``spectral projectors" definition.

\subsection{Topological susceptibility and scale setting}
\label{sec:topological_susceptibilty}

\begin{figure}[H]
\begin{minipage}{0.48\linewidth}
    \includegraphics[width=1.\textwidth]{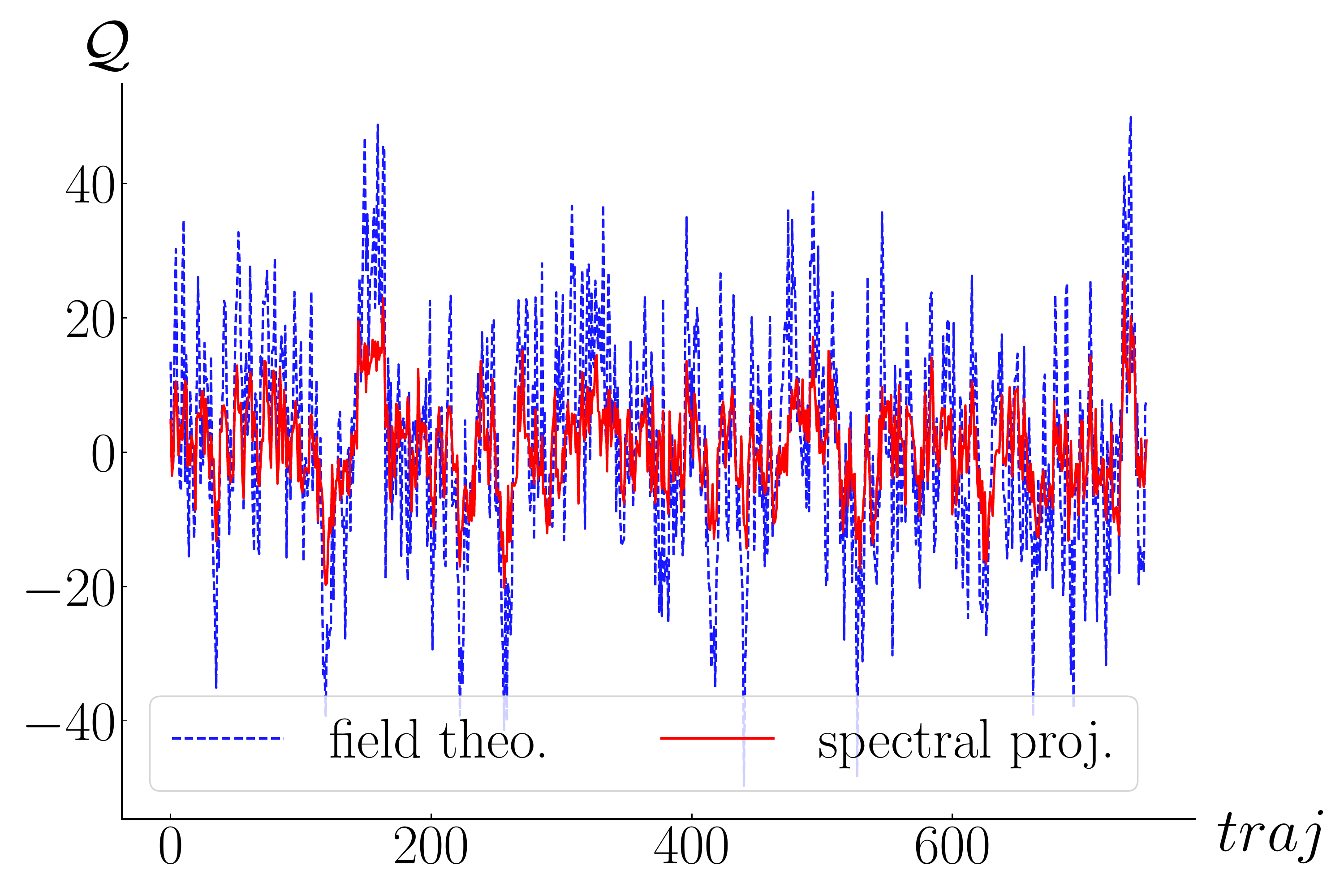}
\end{minipage}%
\hspace*{\fill}
\begin{minipage}{0.48\linewidth}
    \includegraphics[width=1.\textwidth]{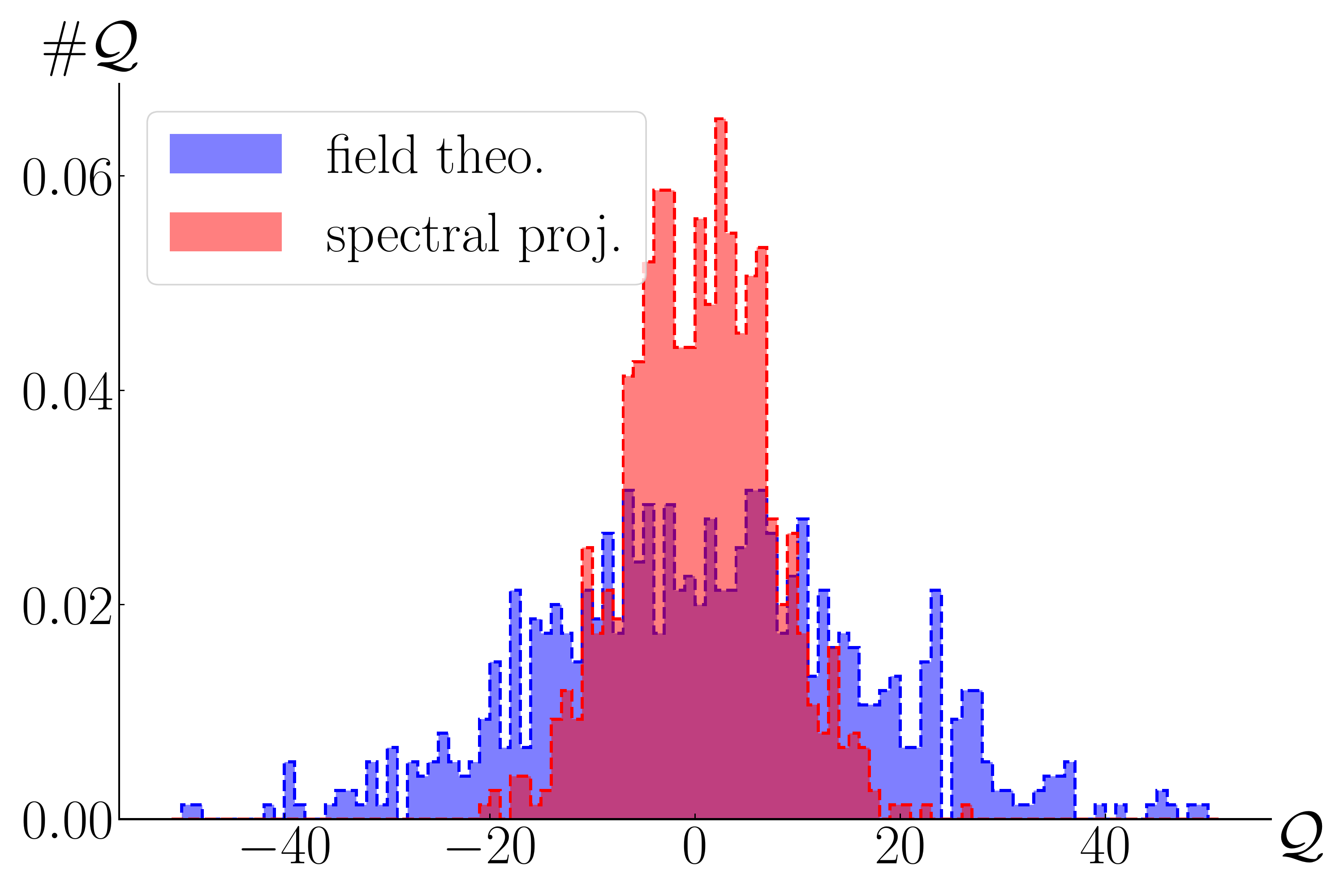}
\end{minipage}
\captionof{figure}{Monte Carlo history of ${\cal Q}$ (left panel), and resulting histogram (right panel) computed on 750 configurations (selected every fourth trajectory), using the  gluonic \(\tau_{\rm flow}=3.5\) (blue) and fermionic with \(M_{\rm thr} = 64.98\:{\rm MeV}\) (red) definitions. }
\label{fig:topocharge}
\end{figure}

Before presenting results on the $CP$-odd form factors, we investigate the topological properties of the cB211.72.64 gauge ensemble.  
In the left panel of Fig.~\ref{fig:topocharge} we show the Monte Carlo (MC) history of the topological charge using the two definitions. One can observe that the autocorrelation time of ${\cal Q}$ is not critical, i.e. no topological freezing occurs at this lattice spacing, and the configurations explore several topological sectors. Moreover, there is a correlation of \(72(2)\%\) between results obtained using the gluonic and fermionic definitions. These results are computed at  \(\tau_{\rm flow}=3.5\), and \(M_{\rm thr} = 64.98\:{\rm MeV}\) but similar observations hold also for other values of the smoothing  and cut-off scales. 

\begin{figure}[H]
\begin{minipage}{0.48\linewidth}
    \includegraphics[width=1.\textwidth]{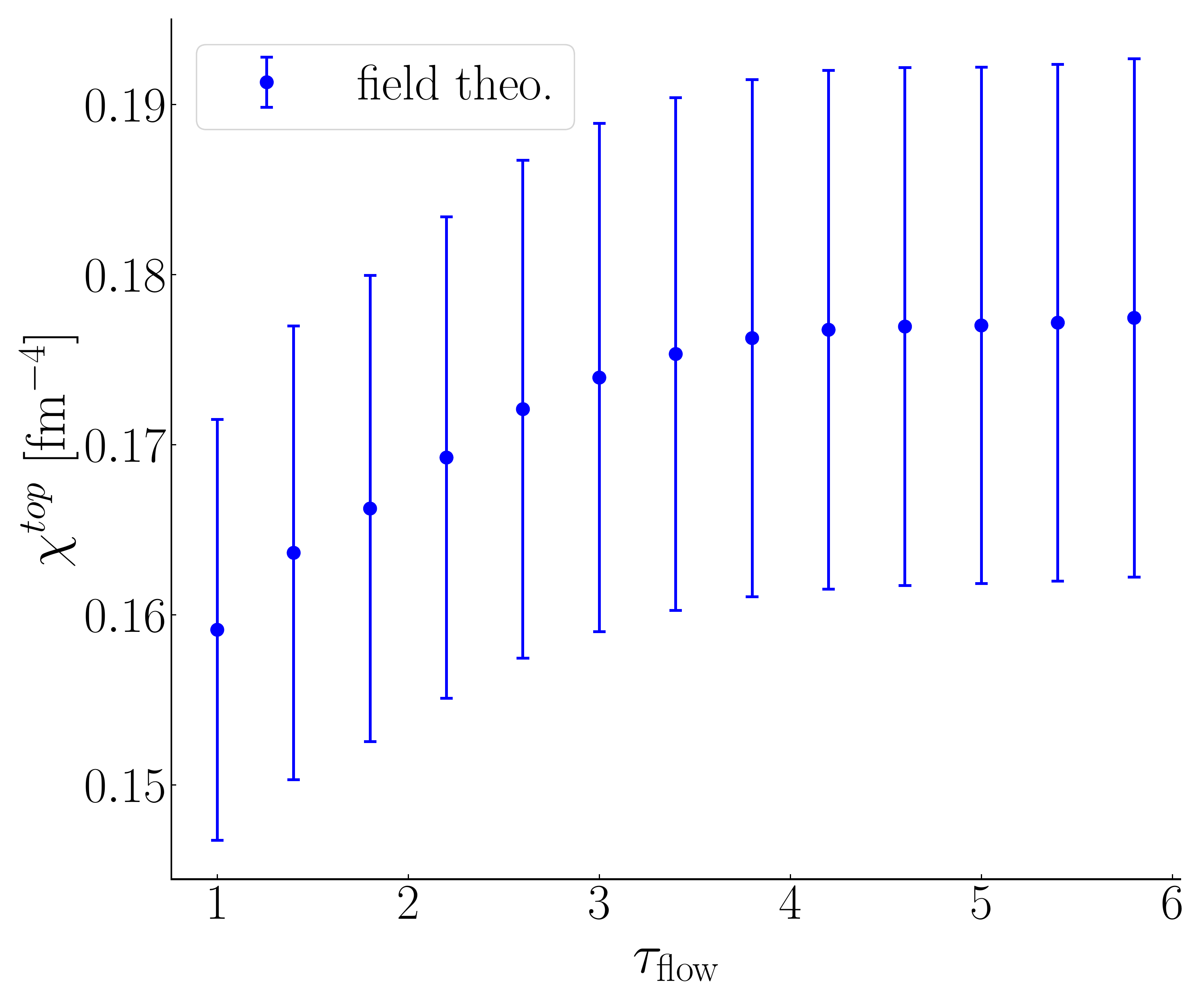}
\end{minipage}%
\hspace*{\fill}
\begin{minipage}{0.48\linewidth}
    \includegraphics[width=1.\textwidth]{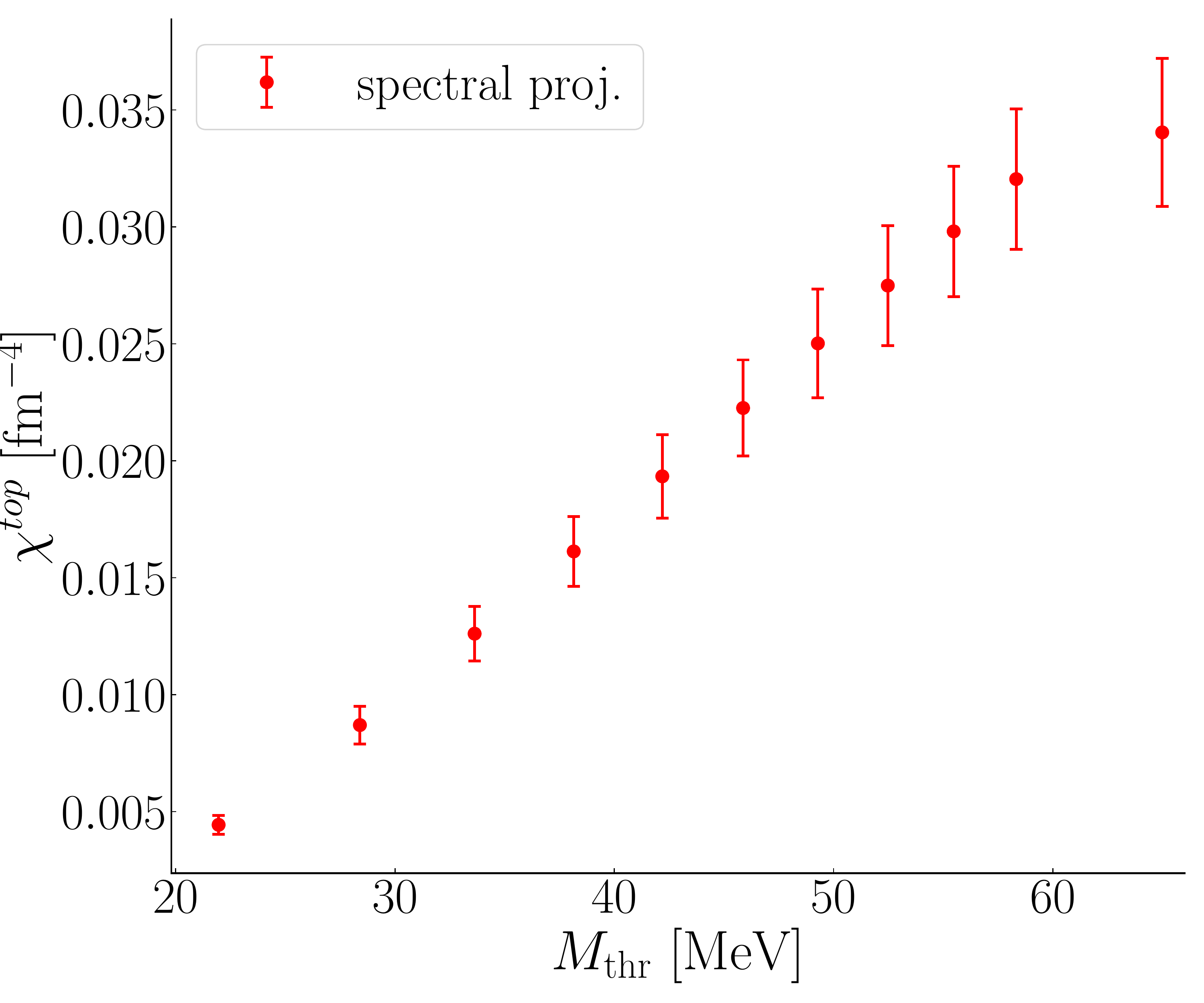}
\end{minipage}
\vspace*{-10pt}
\captionof{figure}{Topological susceptibility with the two definitions of \(\cal Q\) using the gluonic definition (left) as a function of the time flow and the fermionic  (right). }
\label{fig:susc_fg}
\end{figure}
In the right panel of Fig.~\ref{fig:topocharge}, histograms of ${\cal Q}$ are compared. Both definitions lead to a Gaussian-like distribution with a mean value of the topological charge compatible with zero within errors with, however, different widths. In particular, the value of the topological charge extracted via spectral projectors shows less fluctuations and, thus, exhibits a narrow distribution. This impacts the values of the topological susceptibility defined as \(\braket{\mathcal{Q}^2}/V\) that differs between the two definitions, as can be seen from Fig.~\ref{fig:susc_fg}. This mismatch is actually expected, due to the fact that the spectral projectors definition shows milder cut-off effects with respect to \(\braket{\mathcal{Q}^2}/V\) computed via the field theoretical definition. This was already observed in Ref.~\cite{Alexandrou:2017bzk}.
Another observation that emerges from Fig.~\ref{fig:susc_fg} is that, while the susceptibility computed using  the field theoretical definition is essentially flat as a function of the time flow, the one computed using spectral projectors depends strongly on the threshold \(M_{\rm thr}\).
The absence of a plateau for the window of eigenvalues computed makes it harder to properly justify a good choice of \(M_{\rm thr}\). In principle, there is no reason to expect to obtain a plateau for a large value of the renormalized cut-off. Actually, the results from Ref.~\cite{Alexandrou:2017hqw} suggest that for the particular lattice spacing we cannot observe a plateau for  cut-offs up to \(180\) MeV.

However, the fact that we do not observe a plateau is not a problem \textit{di per se}. Indeed, the particular choice of \(M_{\rm thr}\) becomes irrelevant once one takes the continuum limit, as long as the renormalized value of the threshold in physical units is kept fixed. 
Since in this work we use one gauge ensemble, the question that arises is what are the finite-size effects that result from taking a threshold that is not in the plateau region.
\begin{figure}[H]
    \centering
    \includegraphics[width=0.65\textwidth]{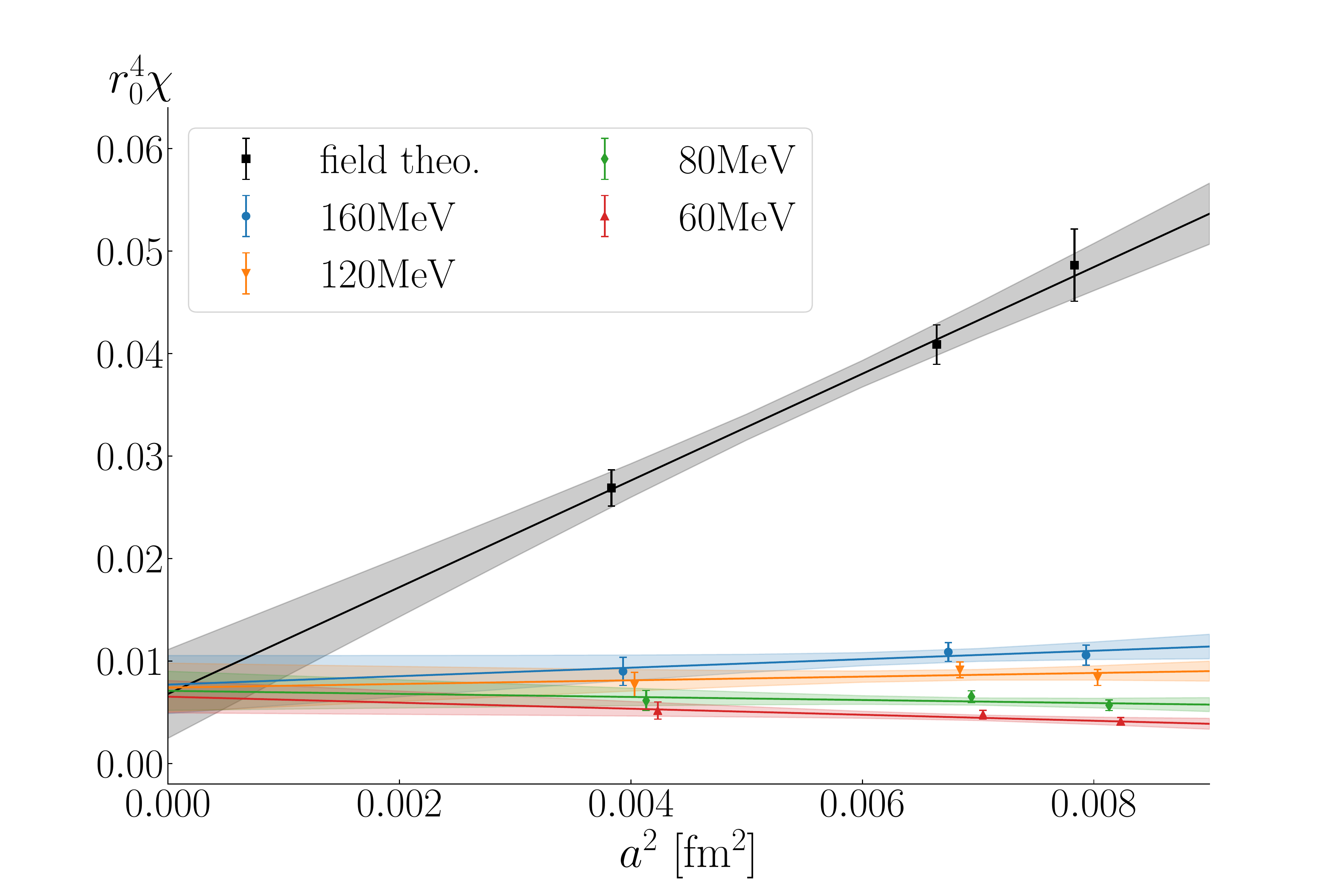}
    \captionof{figure}{Continuum extrapolation of topological susceptibility taken from Ref. \cite{Alexandrou:2017bzk}, with the adjoint of points at \(M_{\rm thr}=60\:{\rm MeV}\). Lattice spacing used in that work corresponds to \(a/r_0=0.035\). Finite-size effects coming from different choices of \(M_{\rm thr}\) are negligible in respect to the ones that affect the field theoretical definition. }
    \label{fig:extrap}
\end{figure}
To give a qualitative answer we use the results obtained in the work of  Ref.~\cite{Alexandrou:2017bzk}. In Fig.~\ref{fig:extrap}, we show the continuum extrapolation of the topological susceptibility for  three values of the threshold used in Ref.~\cite{Alexandrou:2017bzk}, plus a fourth value, at \(M_{\rm thr}=60{\rm MeV}\). What can be seen is that, even if the slope of the continuum limit increases when we decrease the threshold, it is still milder than the slope of the susceptibility computed via the gluonic definition. Thus, the error in taking the value at finite \(a\) instead of its continuum extrapolation, is negligible with respect to the one coming from cut-off effects of the gluonic definition.
Therefore, based on this observation, we use the maximum value of the threshold accessible with the current number of eigenvalues computed, i.e. \(M_{\rm thr}=64.98\:{\rm MeV}\). We can reasonably expect that similar considerations also hold for the quantities entering the determination of the nEDM, i.e. the two- and three-point function correlated with \(\mathcal{Q}\). Their dependence on the smoothing scale (\(\tau_{\rm flow}\) and cut-off  \(M_{\rm thr}\)) is investigated in the following section. 


\section{Results}
\label{sec:results}

\subsection{Nucleon mixing angle}
\begin{figure}[H]
\begin{minipage}{0.48\linewidth}
    \includegraphics[width=1.\textwidth]{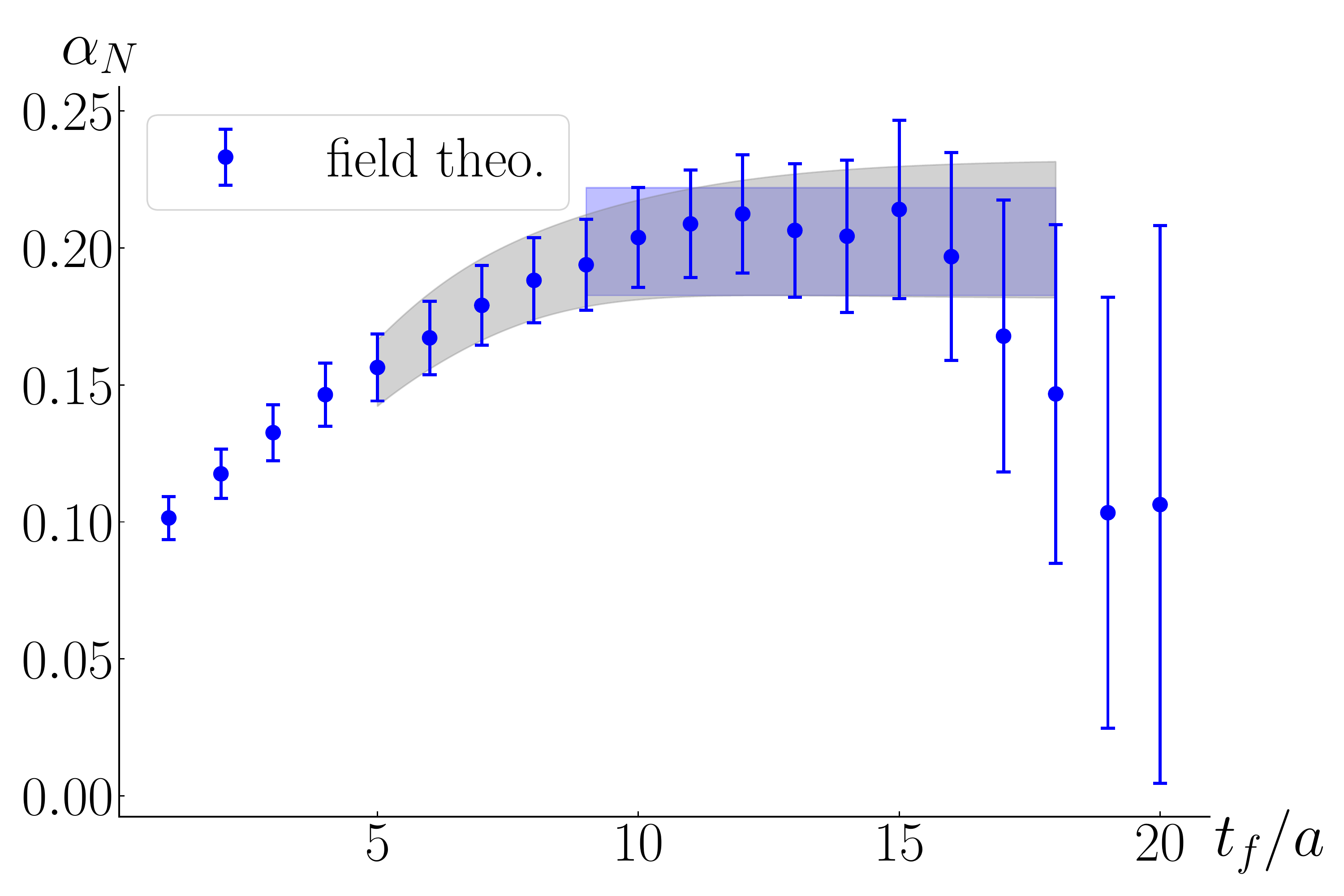}
\end{minipage}%
\hspace*{\fill}
\begin{minipage}{0.48\linewidth}
    \includegraphics[width=1.\textwidth]{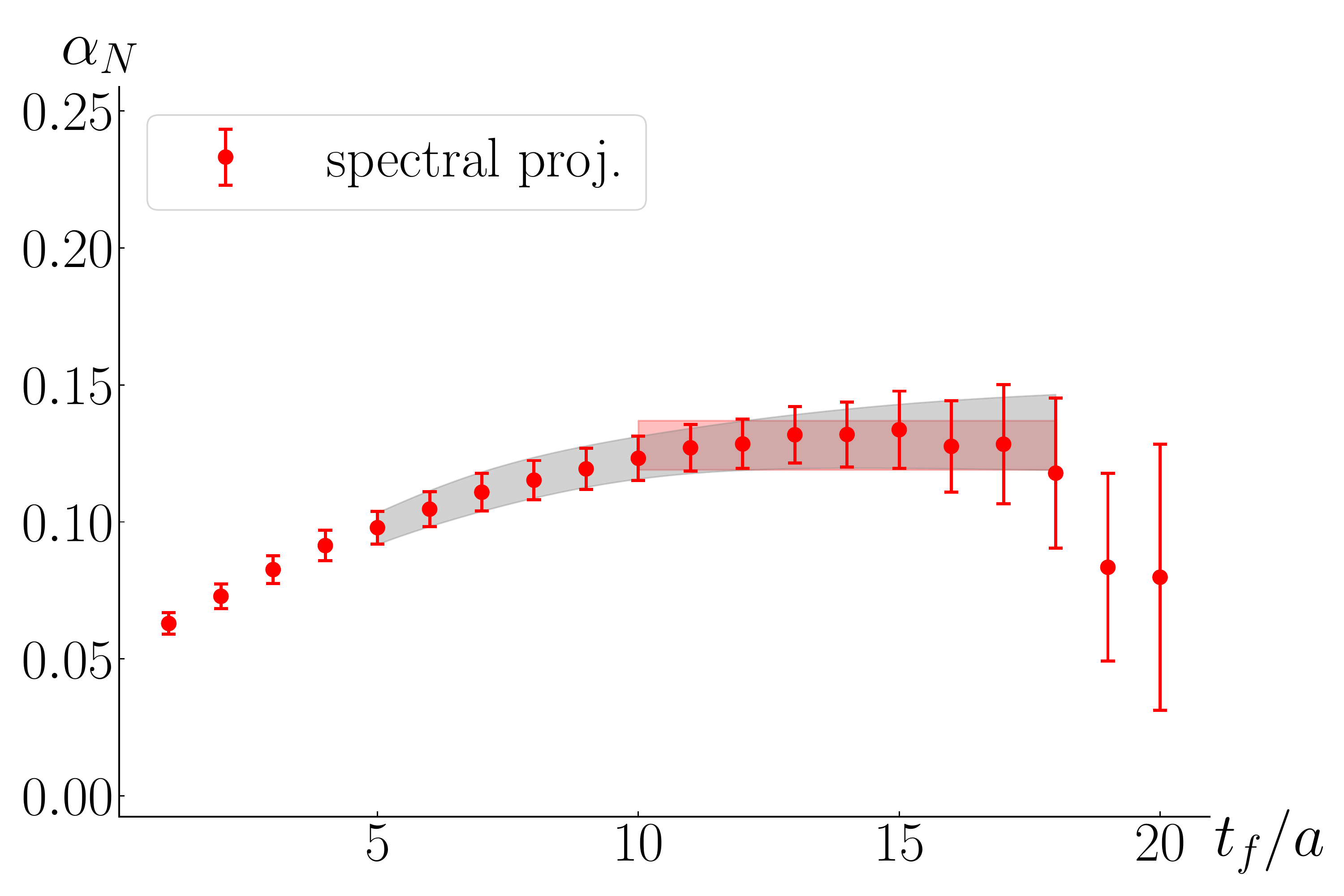}
\end{minipage}
\vspace*{-10pt}
\captionof{figure}{Value of the ratio in Eq.~\eqref{eq:extract_alfa}, as a function of $t_f/a$, using the two definitions of the topological charge, gluonic definition of Eq.~\eqref{eq:Q_fieldteo_def} (left) and fermionic definition of Eq.~\eqref{eq:fQ_ren} (right). With the blue (left) and red (right) bands we show the result of a constant fit within the plateau. With the grey band we show the corresponding fits when using for the fit a constant plus an exponential term which takes into account the first excited state.}
\label{fig:alfa_angle_tf}
\end{figure}

For the determination of the $CP$-odd form factor \(F_3(Q^2)\), one requires the knowledge of the mixing angle \(\alpha_N\). We extract it from the following ratio of two-point functions at zero-momentum
\begin{equation}
    \alpha_N = \lim_{t_f\rightarrow\infty} \frac{G_{2pt,\mathcal{Q}}(\Gamma_5,\vec{0},t_f)}{G_{2pt}(\Gamma_0,\vec{0},t_f)}\,.
    \label{eq:extract_alfa}
\end{equation}
 We take  $t_i=0$ and thus suppress the dependence on \(t_i\). We seek for an interval for which the ratio becomes time-independent as a function of $t_f$ (plateau region). The time evolution of the ratio of Eq. \eqref{eq:extract_alfa} is illustrated in Fig. \ref{fig:alfa_angle_tf}  as a function of \(t_f/a\). We use both the gluonic (left panel) and the fermionic (right panel) definitions for the topological charge that enters in the computation of \(\alpha_N\). In particular, the one in the left panel is computed at timeflow \(\tau_{\rm flow}=3.5\), while for the fermionic one, we used a cut-off of \(M_{\rm thr}=64.98\) MeV.

We identify a plateau region in the range \(t_{f}/a \in [9,18]\) and \(t_{f}/a \in [10,18]\) for the  gluonic and fermionic definitions, respectively. Fitting to a constant, we find \(\alpha_N=0.202(20)(4)\) for the gluonic topological charge and \(\alpha_N=0.128(9)(3)\) for the fermionic one. Errors are respectively the statistical and the systematic coming from the choice of the plateau range. The latter is computed by varying the initial time slice  in the range \([8,12]\) and the final time slice in the interval \([17,20]\), and taking the largest difference between mean values. For both definitions of \(\cal Q\), it is negligible if compared to the statistical uncertainty. To account for a residual time-dependence we consider the contribution of the first excited state too. As shown in Fig.~\ref{fig:alfa_angle_tf}, both Ans\"atze  give compatible results and further validate the choice of the plateau region.

\begin{figure}[H]
\begin{minipage}{0.48\linewidth}
    \includegraphics[width=1.\textwidth]{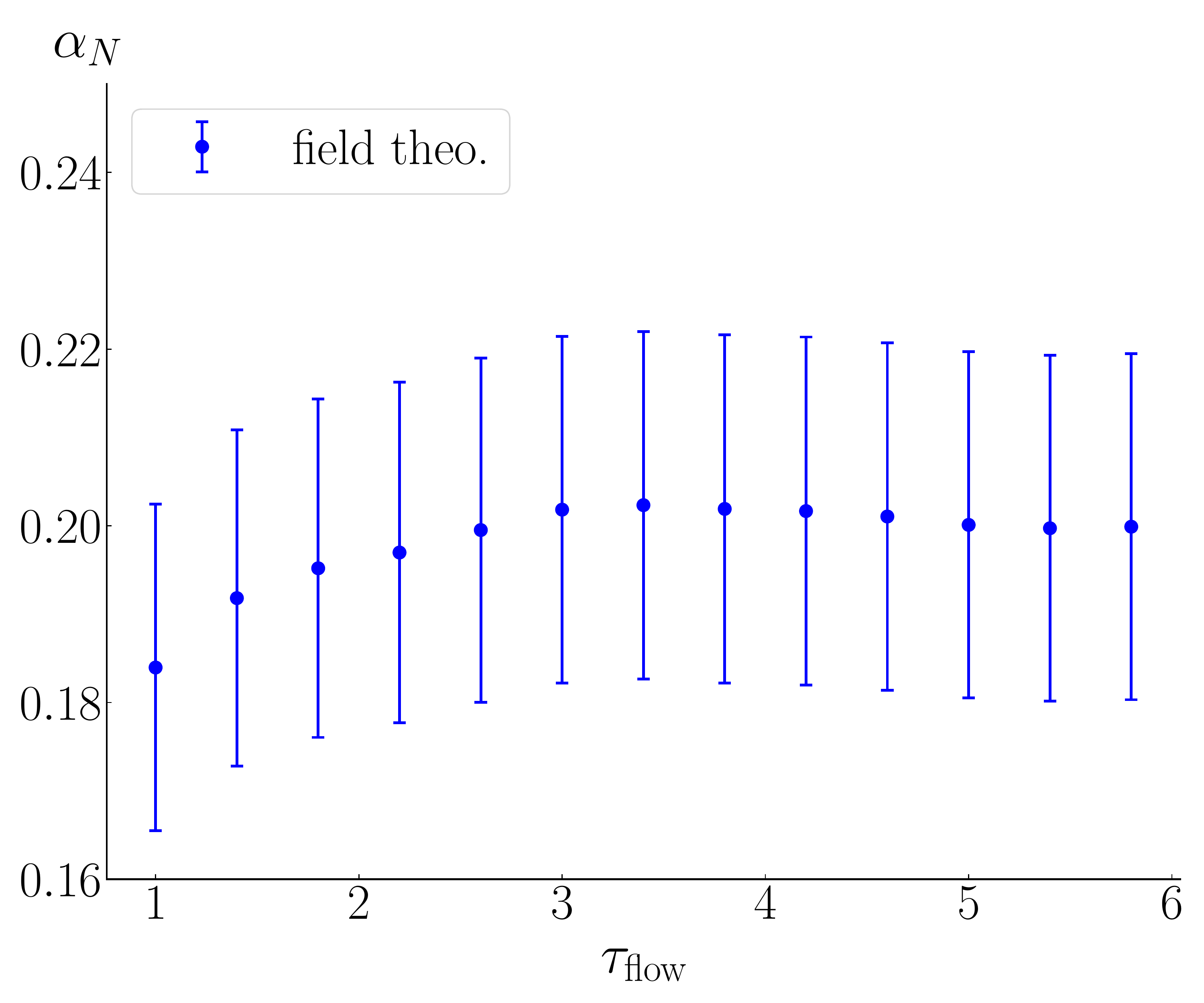}
\end{minipage}%
\hspace*{\fill}
\begin{minipage}{0.48\linewidth}
    \includegraphics[width=1.\textwidth]{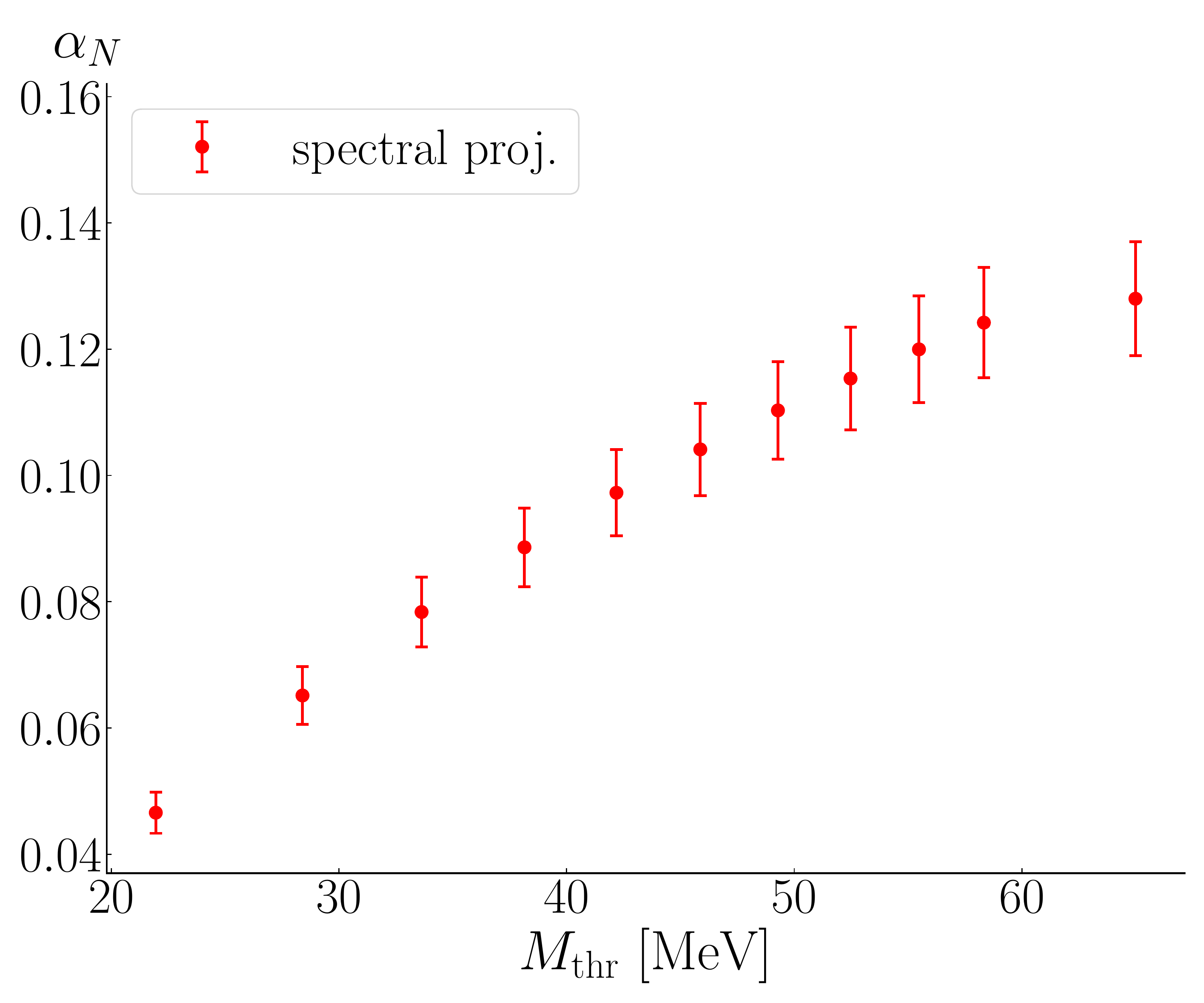}
\end{minipage}
\captionof{figure}{Dependence of the nucleon mixing angle $\alpha_N$ on the smoothing scale $\tau_{\rm flow}$ and cut-off \(M_{\rm thr}\). Left, using the gluonic definition of the topological charge and right, using spectral projectors. In both cases $\alpha_N$ is extracted using a constant fit within the plateau region of the ratio of Eq.~(\ref{eq:extract_alfa}).}
\label{fig:alfa_angle_tc}
\end{figure}

In Fig. \ref{fig:alfa_angle_tc}, we show the dependence of $\alpha_N$ on the smoothing scale of the gluonic definition of the topological charge. Values reported are extracted using a constant fit in the plateau region. As can be seen, from  \(\tau_{\rm flow}=3\), our resulting value of \(\alpha_N\) does not depend on \(\tau_{\rm flow}\). 
In the same figure we also show the dependence of $\alpha_N$ on the threshold \(M_{\rm thr}\) computed using the spectral projectors for the topological charge and a constant fit. This shows a stronger dependence on \(M_{\rm thr}\). Such a behaviour is reminiscent  of the one   exhibited by the topological charge as discussed in Section~\ref{sec:topological_susceptibilty}.

\subsection{Determination of the $CP$-odd form factor $F_3(Q^2)$}
\label{sec:cp_odd_form_factor}
In order to extract the $CP$-odd form factor $F_3(Q^2)$, we construct an appropriate ratio of the relevant three-point function given in Eq.\eqref{eq:G3ptQ_prjcted} and a combination of two-point functions  so that  unknown overlaps and the exponential time dependence cancel in the large time limit of $t_f$ and $t_{\rm ins}$. This ratio is given by
\begin{equation}
    \Pi^{\mu k}_{3pt,\mathcal{Q}}(\vec{q}) \equiv \lim_{t_f,t_{ins}\rightarrow\infty} \frac{G^{\mu}_{3pt,\mathcal{Q}}(\Gamma_k,\vec{q},t_f,t_{ins}) }{ G_{2pt}(\Gamma_0,\vec{0},t_f) } \sqrt{\frac{G_{2pt}(\Gamma_0,\vec{q},t_f-t_{ins}) G_{2pt}(\Gamma_0,\vec{0},t_{ins}) G_{2pt}(\Gamma_0,\vec{0},t_f) }{G_{2pt}(\Gamma_0,\vec{0},t_f-t_{ins})G_{2pt}(\Gamma_0,\vec{q},t_{ins})
    G_{2pt}(\Gamma_0,\vec{q},t_f)}}\,,
    \label{eq:Pi_munu}
\end{equation}%
where \(\vec{p}_f=\vec{0}\).
The form factor \(F_3(Q^2)\) is then extracted from
\begin{equation}
\Pi^{0k}_{3pt,\mathcal{Q}}(\vec{q}) = \frac{i q_k \mathcal{C}}{2m_N} \left(\alpha_N  G_E(Q^2)-\frac{F_3(Q^2)}{2m_N} (E_N+m_N)   \right)\,,
    \label{eq:F3_extract}
\end{equation}
where \(E_N\) is the initial energy of the nucleon, \(\mathcal{C}=\sqrt{(2m_N^2)/(E_N(E_N+m_N))}\) is a kinematic factor and \(G_E(Q^2) = F_1(Q^2) + (q^2/(2m_N^2)) F_2(Q^2)\) is the electric Sachs form factor. $F_1(Q^2)$ and $F_2(Q^2)$ are the Pauli and Dirac form factors. We note that Eq.~(\ref{eq:F3_extract}) derives from Eq.~(55) of Ref.~\cite{Alexandrou:2015spa} with the correction given in Ref. \cite{Abramczyk:2017oxr}. The neutron electric form factor \(G_E(Q^2)\) is extracted from  \(\Pi^{00}_{3pt}\) (see  Eq.~(A4) of Ref.~\cite{Alexandrou:2018sjm})
\be
\Pi^{00}_{3pt}= \mathcal{C} \frac{E_N+m_N}{2m_N}G_E(Q^2)\,.
\label{eq:Ge}
\ee
We note that Eq.~\eqref{eq:F3_extract} is not the only way to extract the \(F_3(Q^2)\) form factor. Similar relations can be written that express the \(\Pi^{kk}_{3pt,\mathcal{Q}}\) and \(\Pi^{j\neq k}_{3pt,\mathcal{Q}}\) components as  linear combinations of \(F_1\), \(F_2\) and \(F_3\). We verified that they lead to worse signal-to-noise ratios, and thus we opt to use Eq.~\eqref{eq:F3_extract} in this work.

\begin{figure}[H]
\begin{minipage}[c]{0.49\linewidth}
\includegraphics[width=0.88\textwidth]{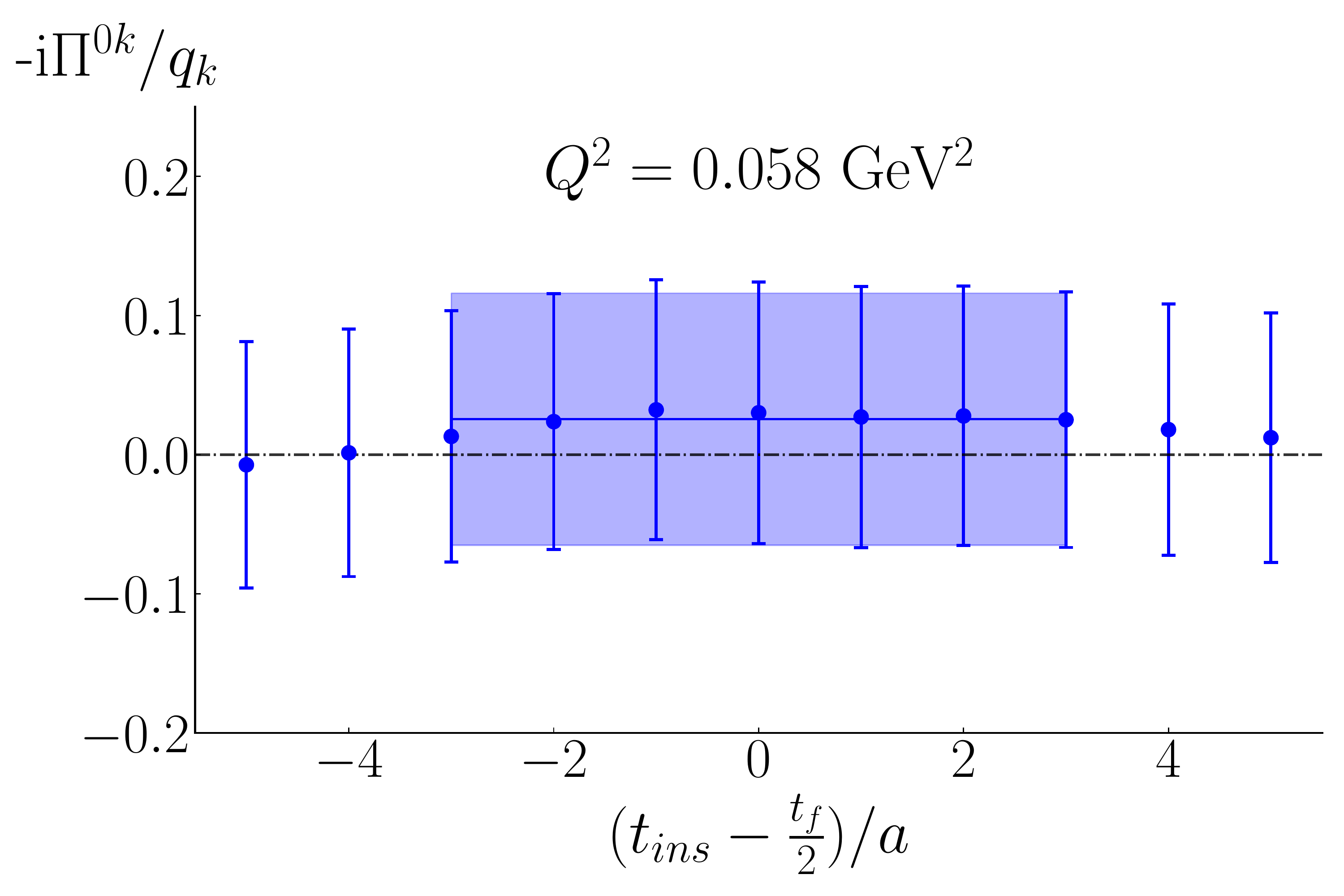}    
\end{minipage}%
\hfill
\begin{minipage}[c]{0.49\linewidth}
\includegraphics[width=0.9\textwidth]{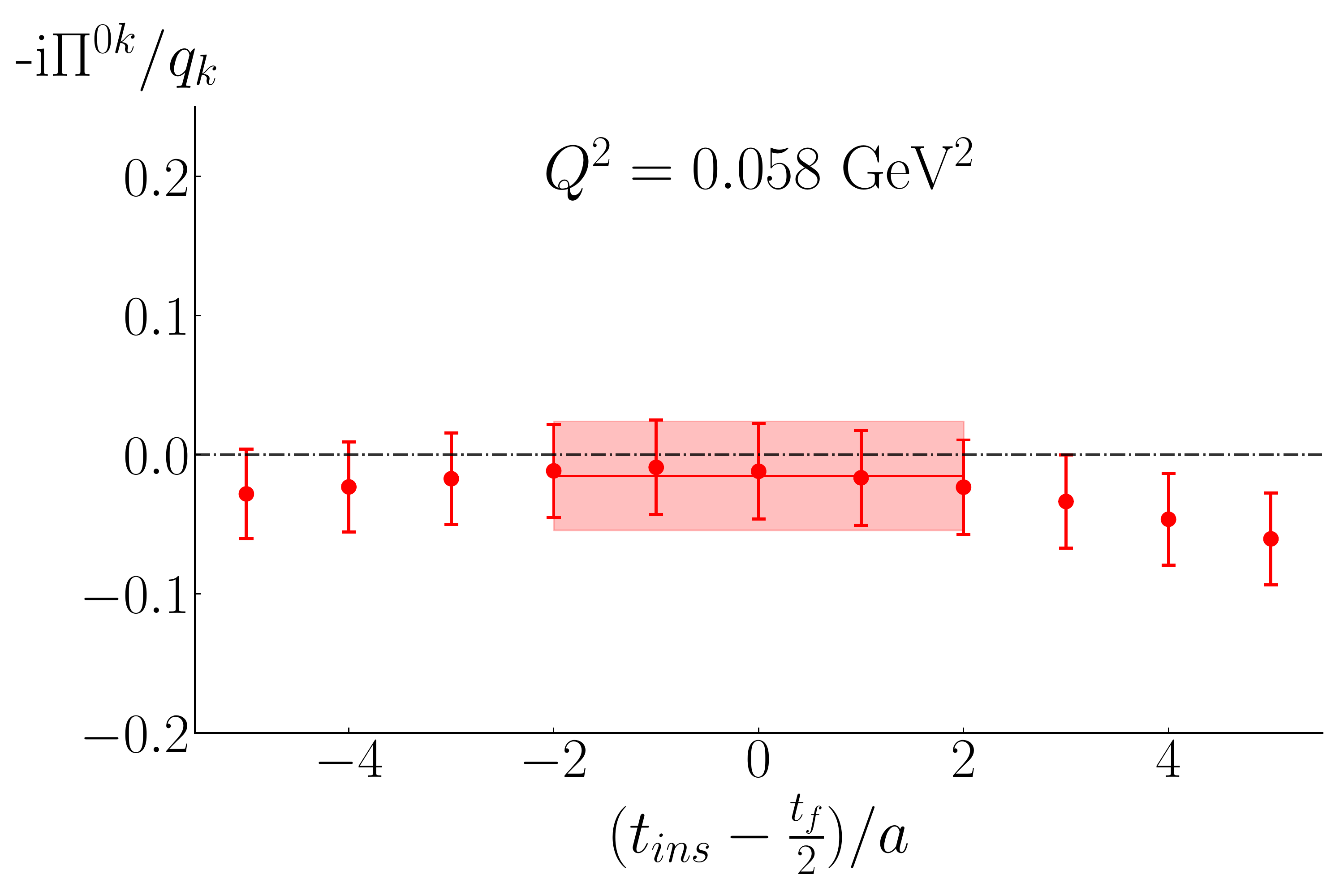}    
\end{minipage}
\begin{minipage}[c]{0.49\linewidth}
\includegraphics[width=0.9\textwidth]{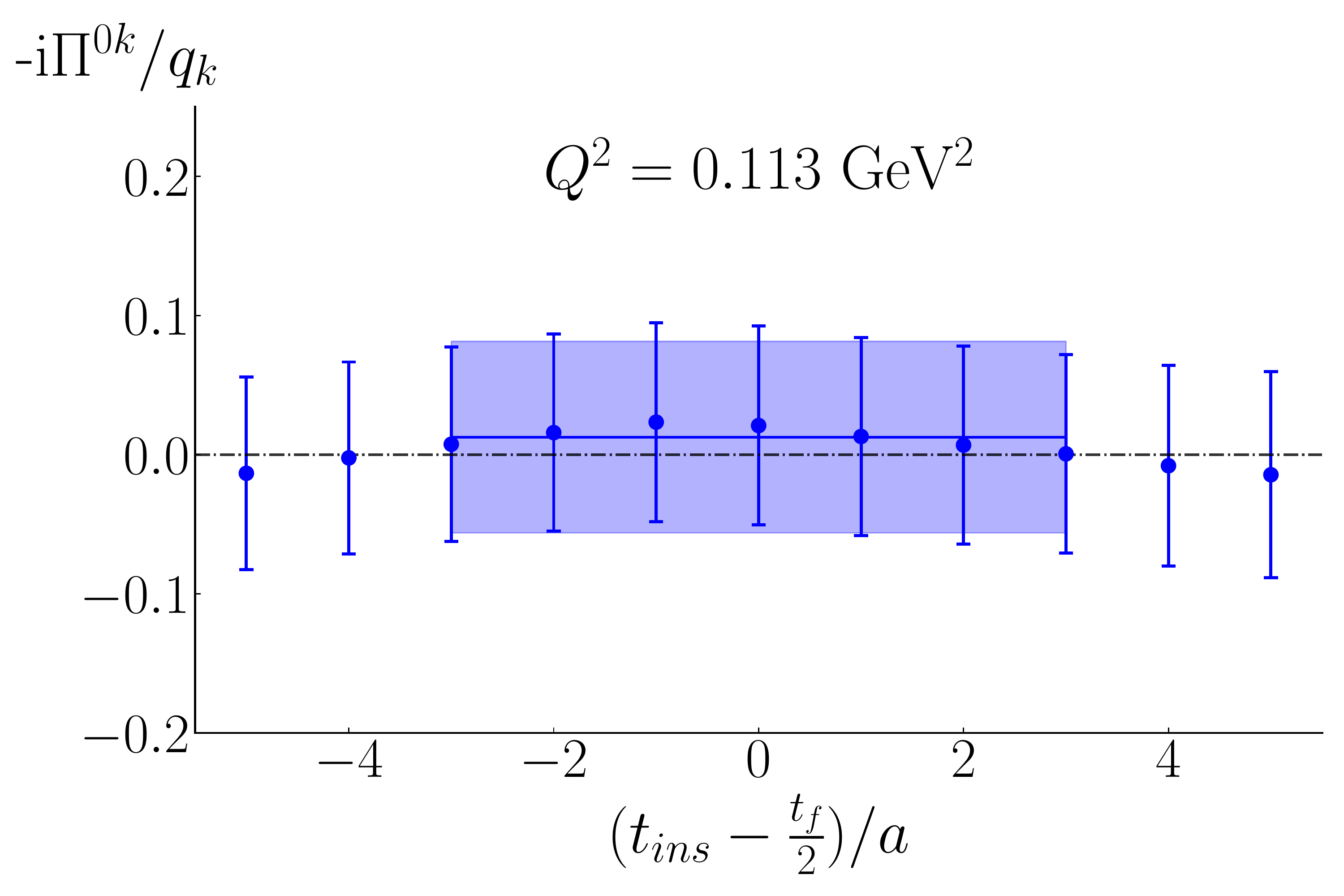}    
\end{minipage}%
\hfill
\begin{minipage}[c]{0.49\linewidth}
\includegraphics[width=0.9\textwidth]{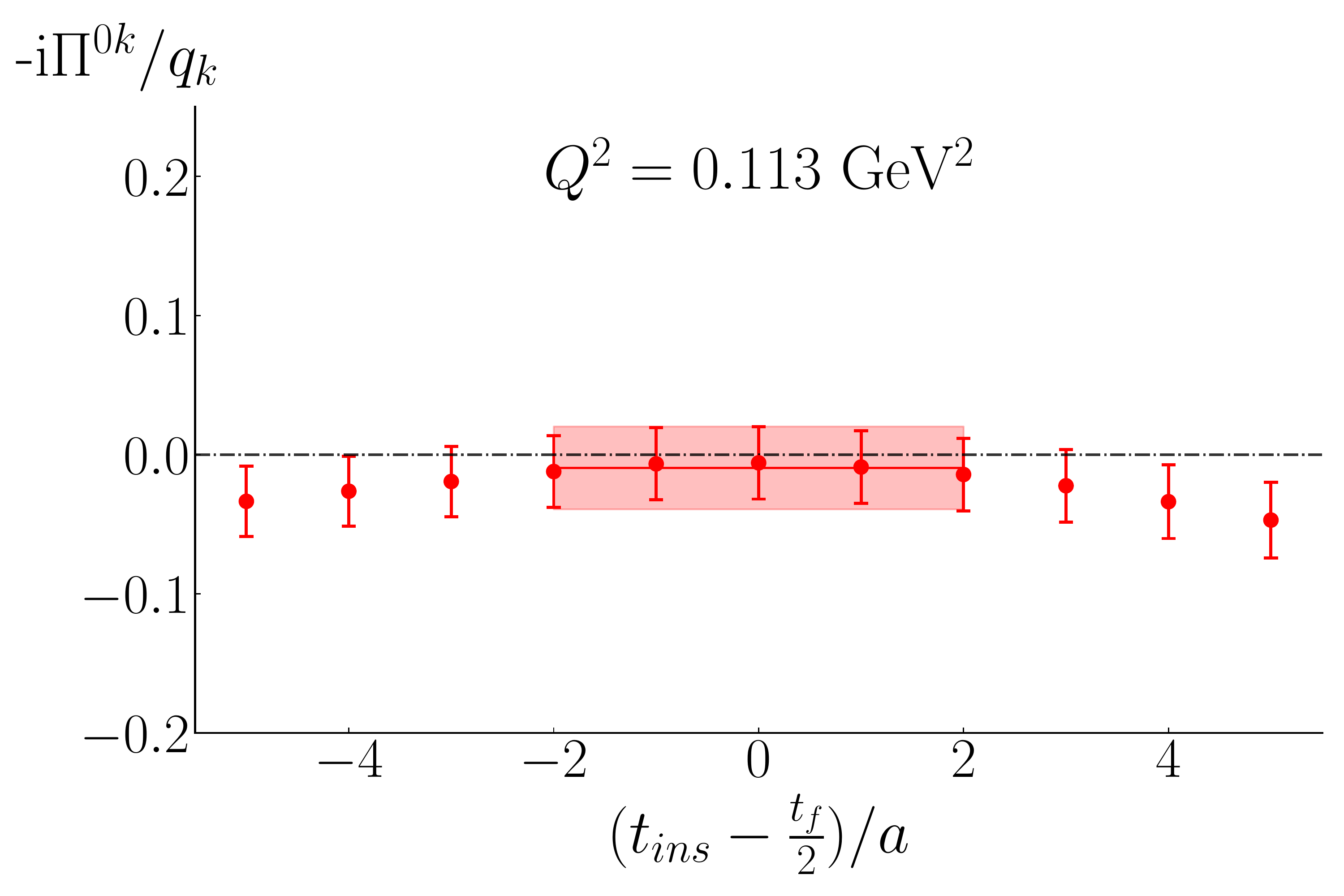}    
\end{minipage}
\begin{minipage}[c]{0.49\linewidth}
\includegraphics[width=0.9\textwidth]{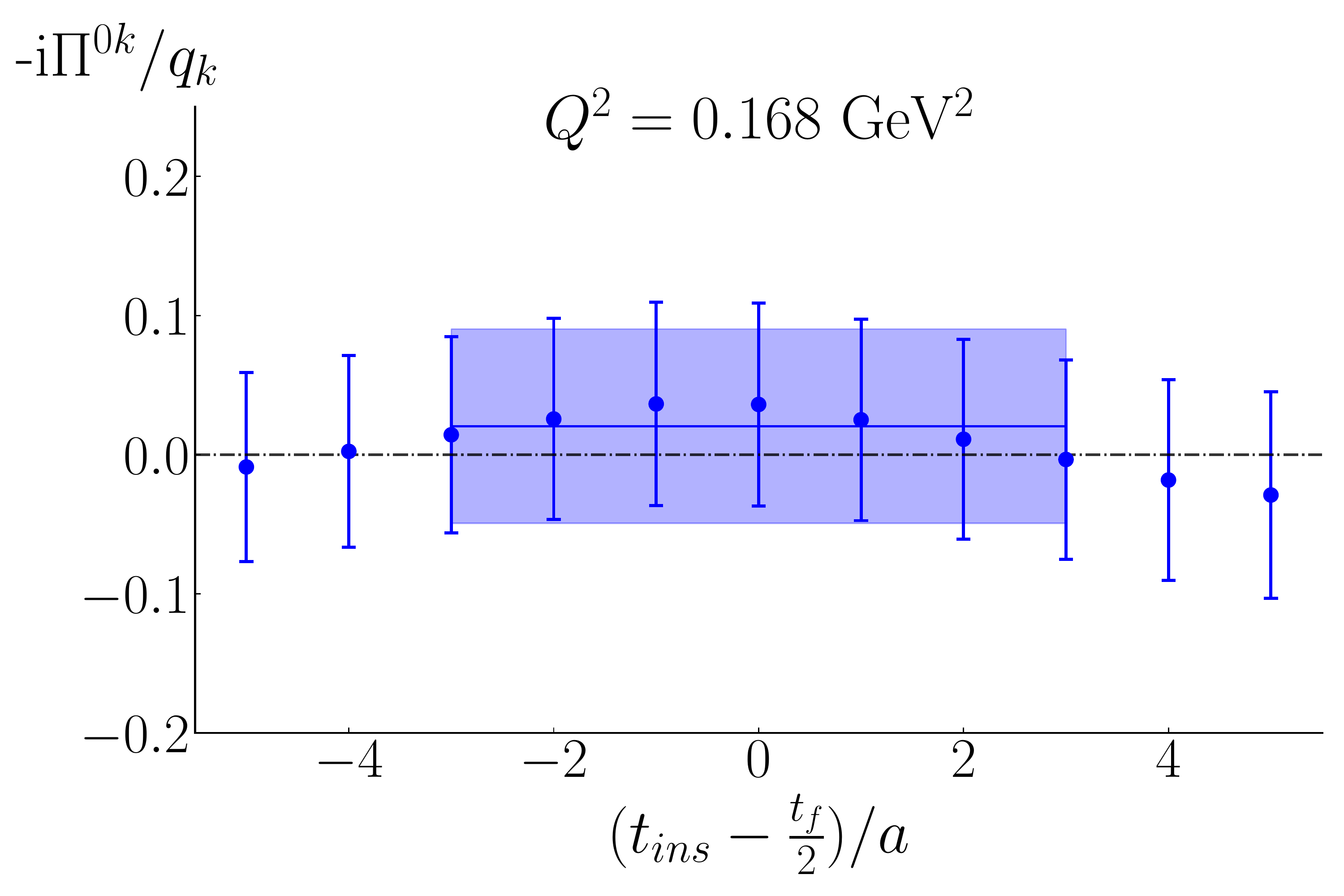}    
\end{minipage}
\hfill
\begin{minipage}[c]{0.49\linewidth}
\includegraphics[width=0.9\textwidth]{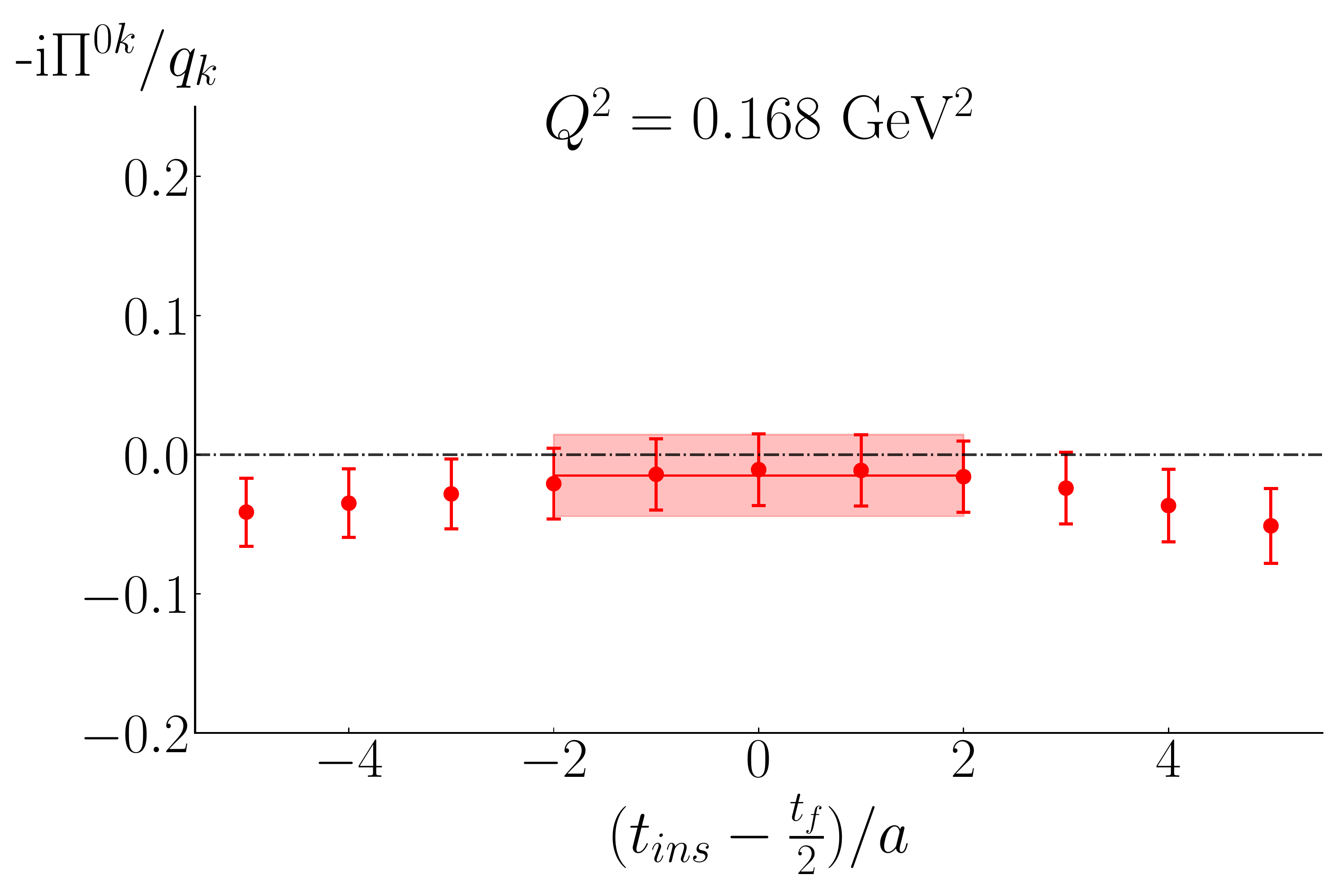}    
\end{minipage}
\captionof{figure}{Ratio of Eq.~\eqref{eq:Pi_munu} as a function of insertion time \(t_{ins}\) at fixed sink-source time separation \(t_f=12a\). The three smallest values of the  momentum transfer squared are shown. In the left column, we show results using  the gluonic  definition of \(\mathcal{Q}\) and  (\(\tau_{\rm flow}=3.5\)), while in the right column, the results are obtained using the spectral projectors for the computation of \(\mathcal{Q}\) and (\(M_{\rm thr}=64.98\) MeV). The bands are  the result of a constant fit in the plateau region excluding symmetrically 3 and 4 time slices for the gluonic (left) and fermionic (right) definition of \(\mathcal{Q}\), respectively.}
\label{fig:tins_plateau}
\end{figure}
\begin{figure}[H]
\begin{minipage}[c]{0.49\linewidth}
\includegraphics[width=1.\textwidth]{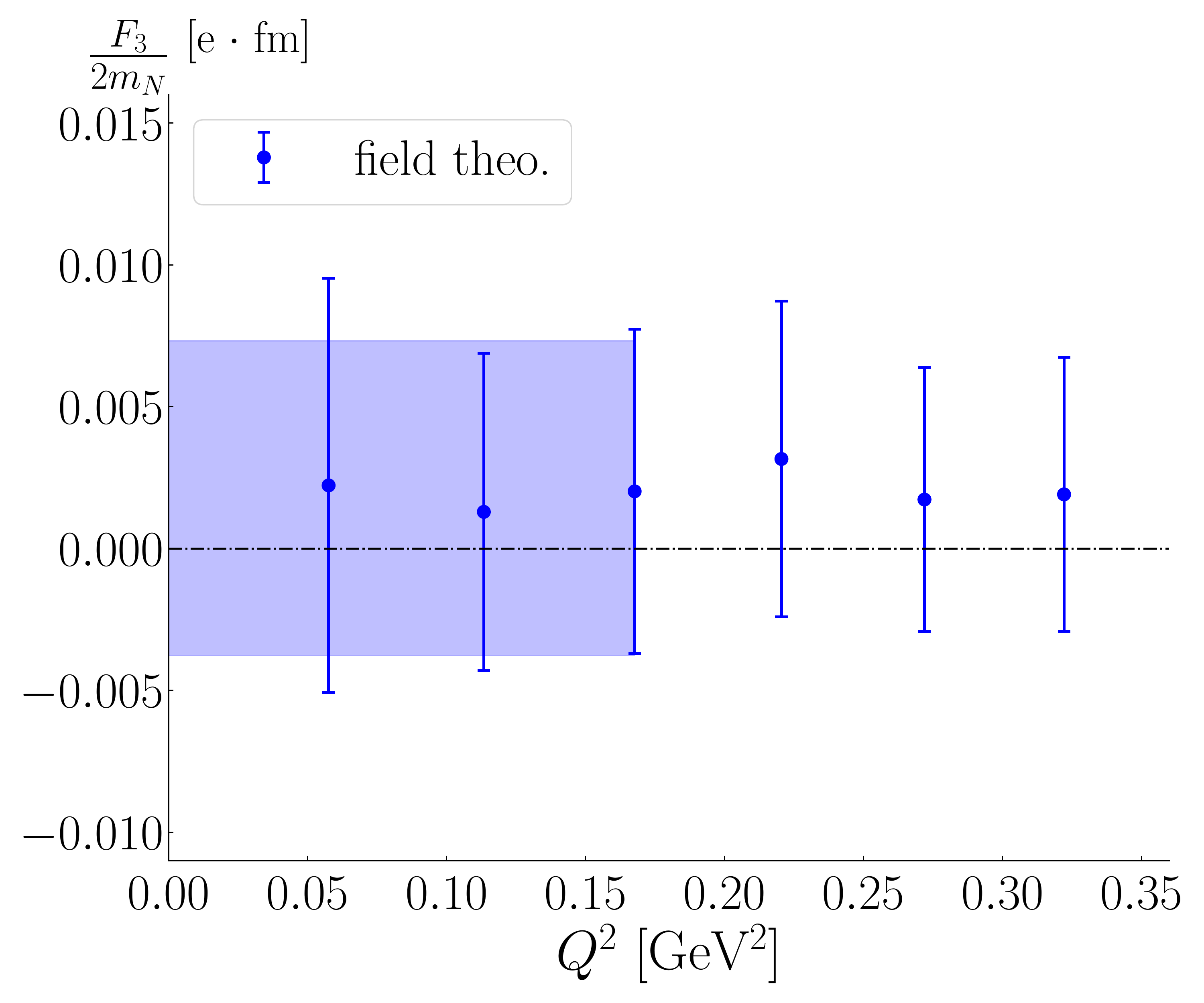}
\end{minipage}%
\hfill
\begin{minipage}[c]{0.49\linewidth}
\includegraphics[width=1.\textwidth]{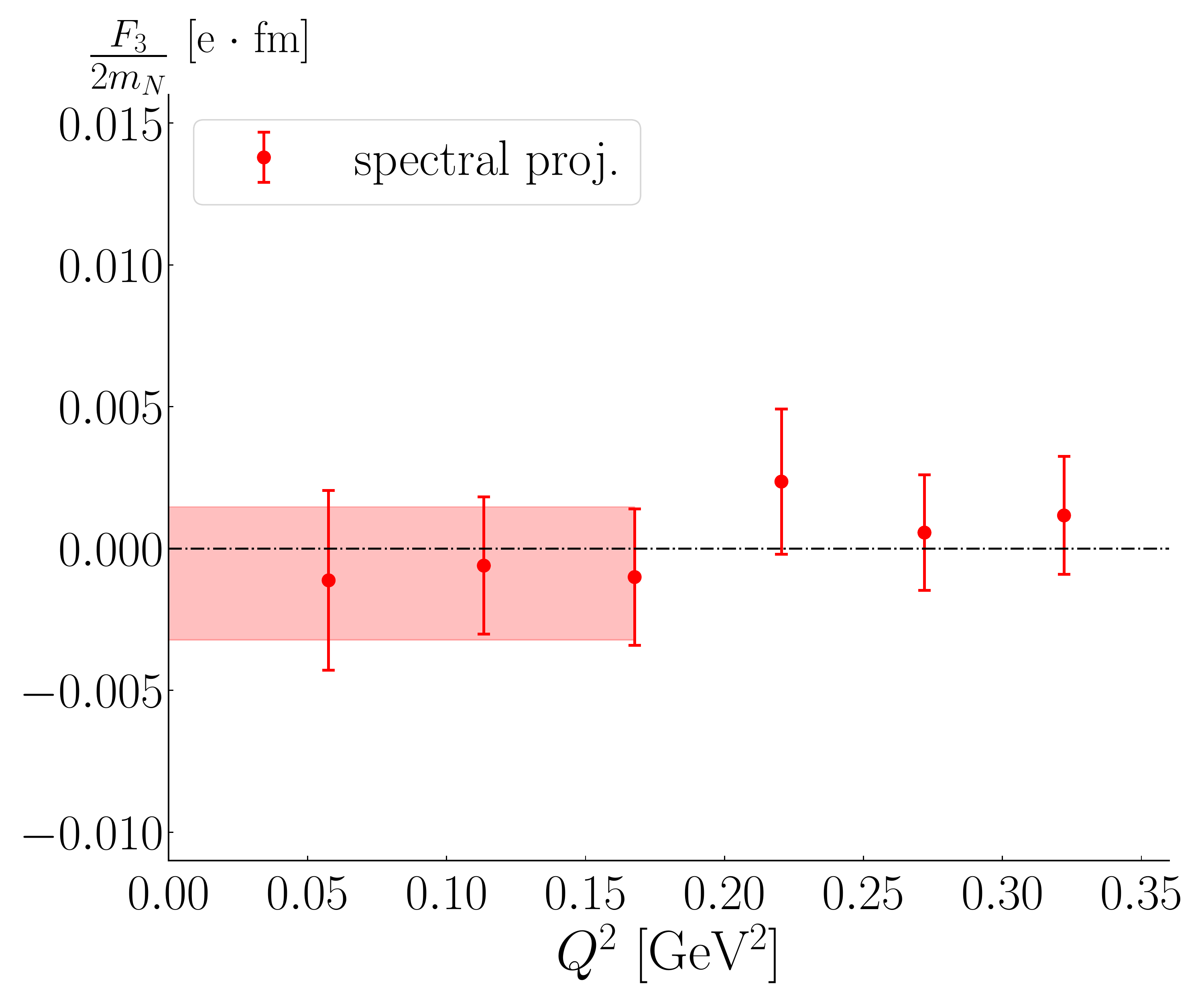}
\end{minipage}
\captionof{figure}{\(F_3(Q^2)\) as a function of \(Q^2\), using a the field theoretical or gluonic definition of the topological charge (left panel) and the fermionic definition based on spectral projectors (right panel). The blue (left) and red (right)  bands represent the weighted average of the values at three smaller values of  \(Q^2\).  }
\label{fig:F3_q0}
\end{figure}

In Fig.~\ref{fig:tins_plateau}, we show the ratio defined in Eq.~\eqref{eq:Pi_munu} as a function of the insertion time \(t_{ins}\) at fixed sink-source time separation \(t_f=12a\), for the smallest three non-zero values of the momentum transfer squared, i.e. \(Q^2 = 0.056\ \text{GeV}^2\), \(Q^2 = 0.111\ \text{GeV}^2\) and \(Q^2 = 0.164\ \text{GeV}^2\). The results shown for \(\Pi^{0k}_{3pt,\mathcal{Q}}\) are averaged among momenta with non-zero \(k\)-component in all \(k\)-directions. Despite the large statistics employed ($\sim$~40000~samples), the errors are large and do not allow to increase further the sink-source time separation.  We fit the ratio in symmetric intervals \([-t_{\rm fit},t_{\rm fit}]\) and vary 
the  fit ranges, taking  \(t_{\rm fit}=2,3,4\). The resulting values are all compatible within our current statistical accuracy.
It is worth to be noting that the results extracted using the fermionic definition of the topological charge show a significant error reduction with respect to the gluonic counterpart. Nevertheless, a zero value cannot be excluded for all three momentum transfers.
This is made even more evident in Fig.~\ref{fig:F3_q0}, where the  values of \(F_3\) at different \(Q^2\) are reported. In order to extract  $F_3(0)$ and thus \(d^{\theta}_N\), we  extrapolate to $Q^2=0$ by considering the weighted average of the values at the three smallest \(Q^2\) values. Other fit forms, such as the dipole Ansatz and the $z$-expansion~\cite{Bhattacharya:2011ah}, or even the momentum elimination technique~\cite{Alexandrou:2020aja}, are not viable with this level of uncertainty.
Our final results are
\begin{align}\label{eq:result_g}   
\text{field~theoretical~or~gluonic definition} \qquad \ \lvert d_N^{\theta}\rvert = 0.0018(56) \: \theta \: {\rm e}\cdot{\rm fm} \,,\\
\text{fermionic definition via spectral~projectors} \qquad \ \lvert d_N^{\theta}\rvert = 0.0009(24)\: \theta \: {\rm e}\cdot{\rm fm}\,.
\label{eq:result_f}   
\end{align}

\begin{figure}[H]
\begin{minipage}[c]{0.33\linewidth}
\centering
\includegraphics[width=1.\textwidth]{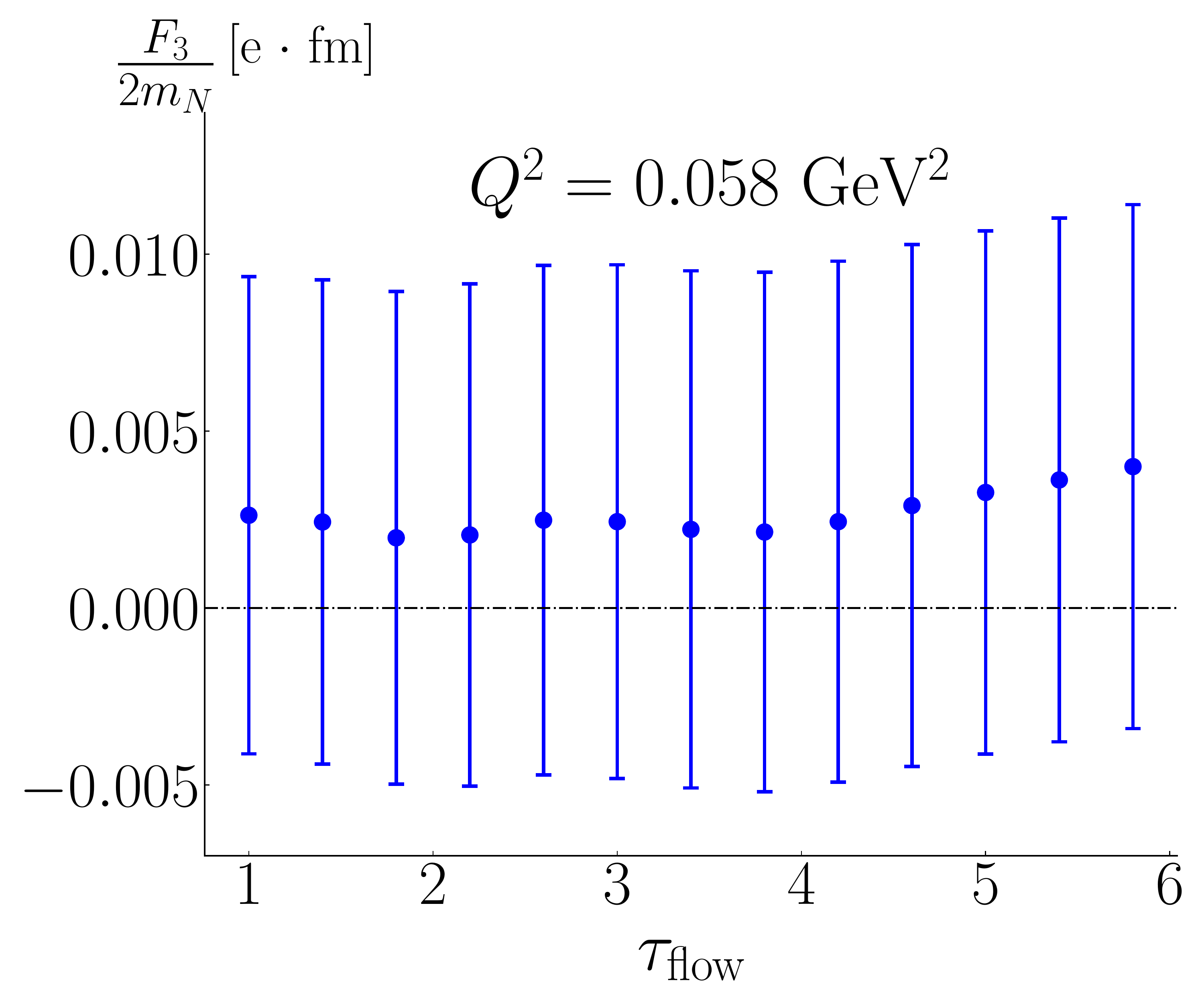}
\end{minipage}%
\hfill
\begin{minipage}[c]{0.33\linewidth}
\centering
\includegraphics[width=1\textwidth]{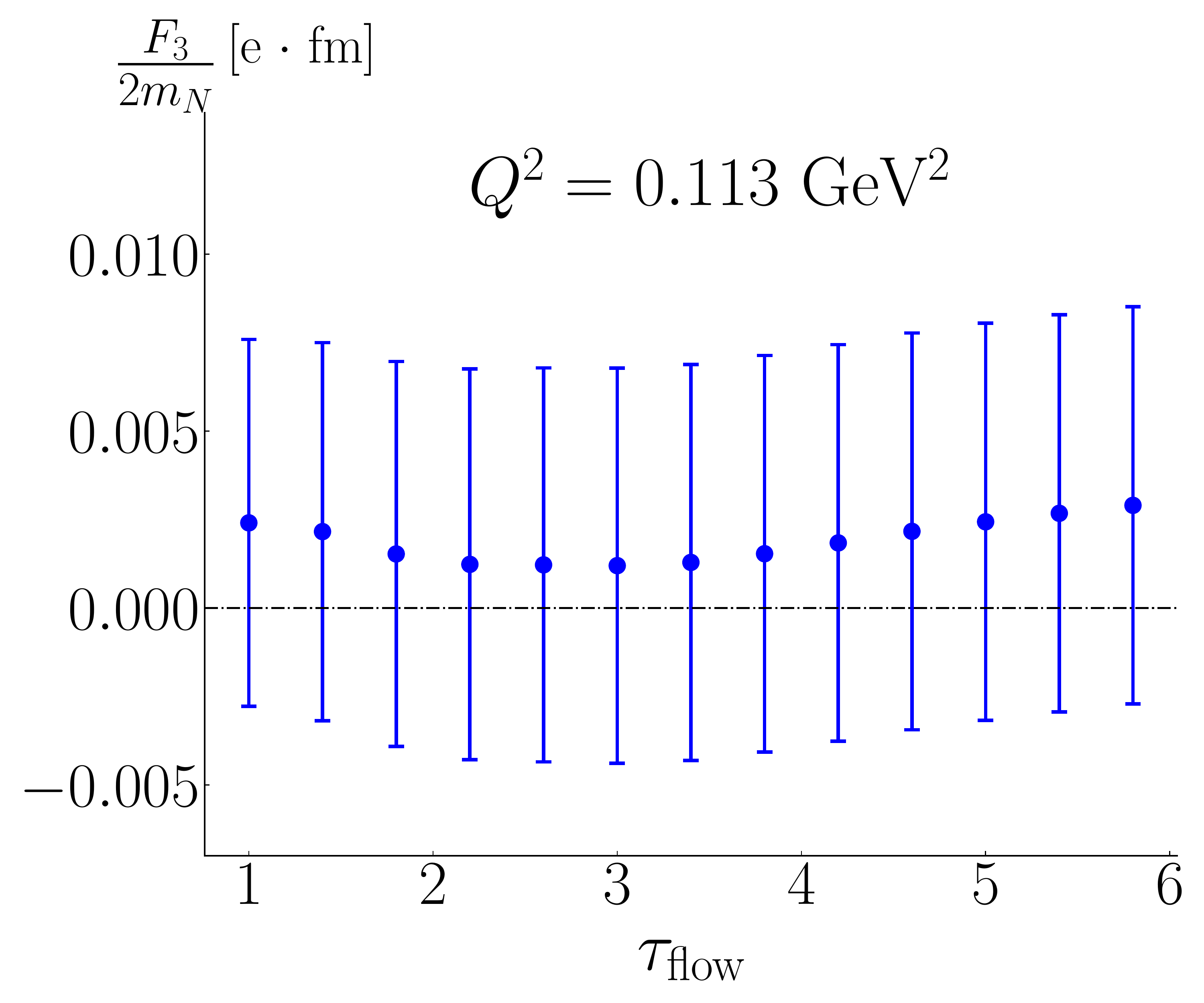}
\end{minipage}%
\hfill
\begin{minipage}[c]{0.33\linewidth}
\centering
\includegraphics[width=1\textwidth]{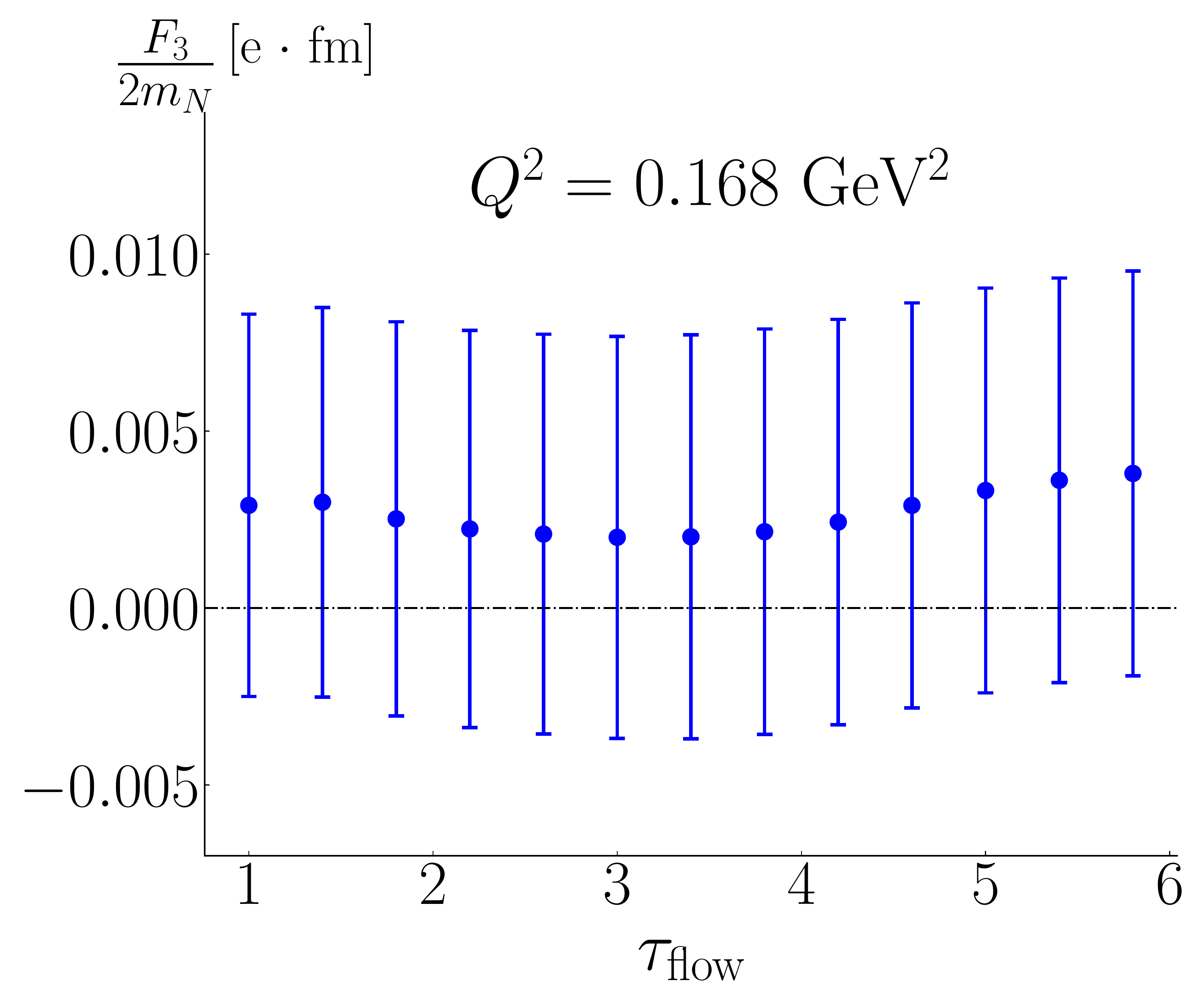}
\end{minipage}
\begin{minipage}[c]{0.33\linewidth}
\centering
\includegraphics[width=1.\textwidth]{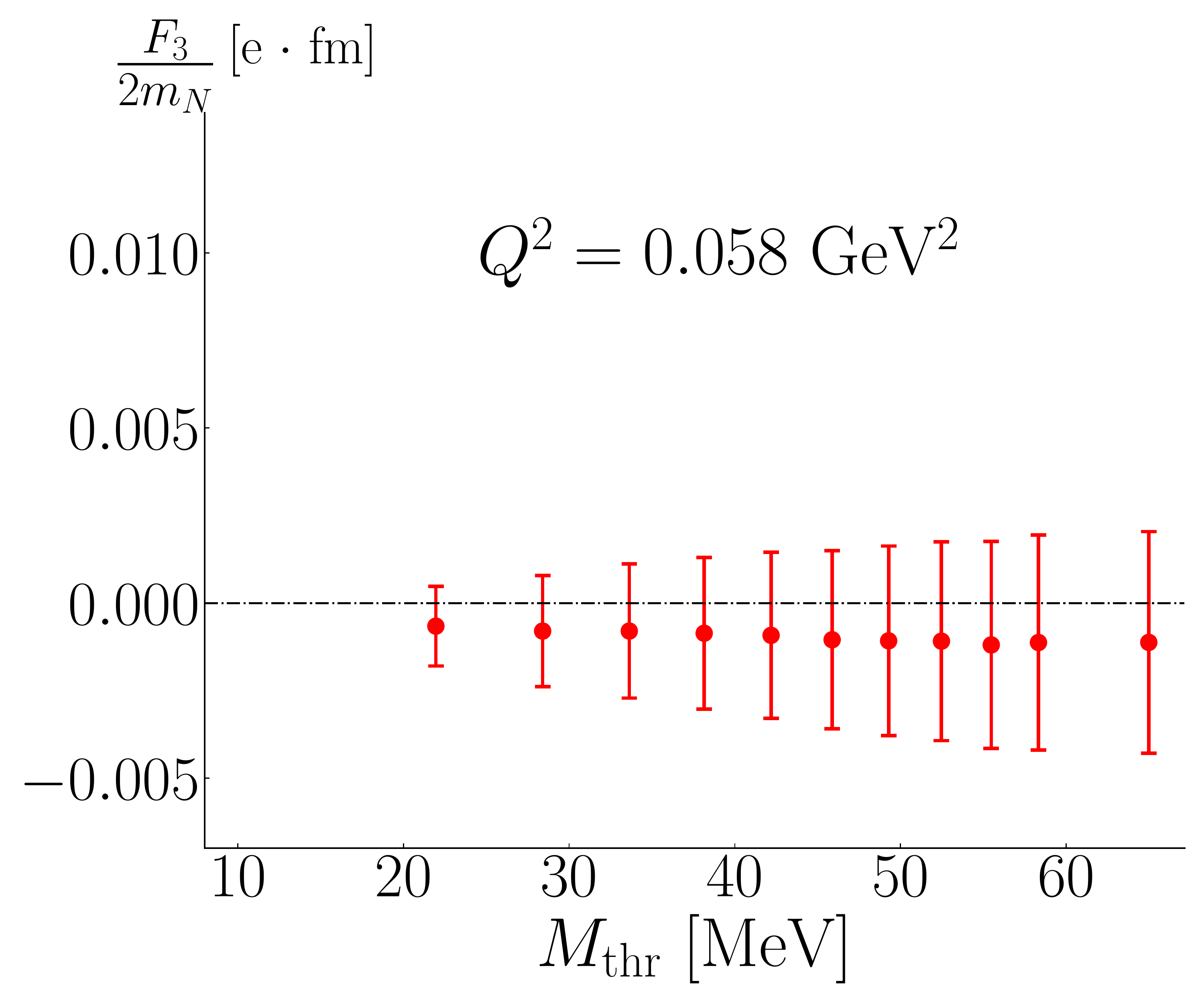}
\end{minipage}%
\hfill
\begin{minipage}[c]{0.33\linewidth}
\centering
\includegraphics[width=1.\textwidth]{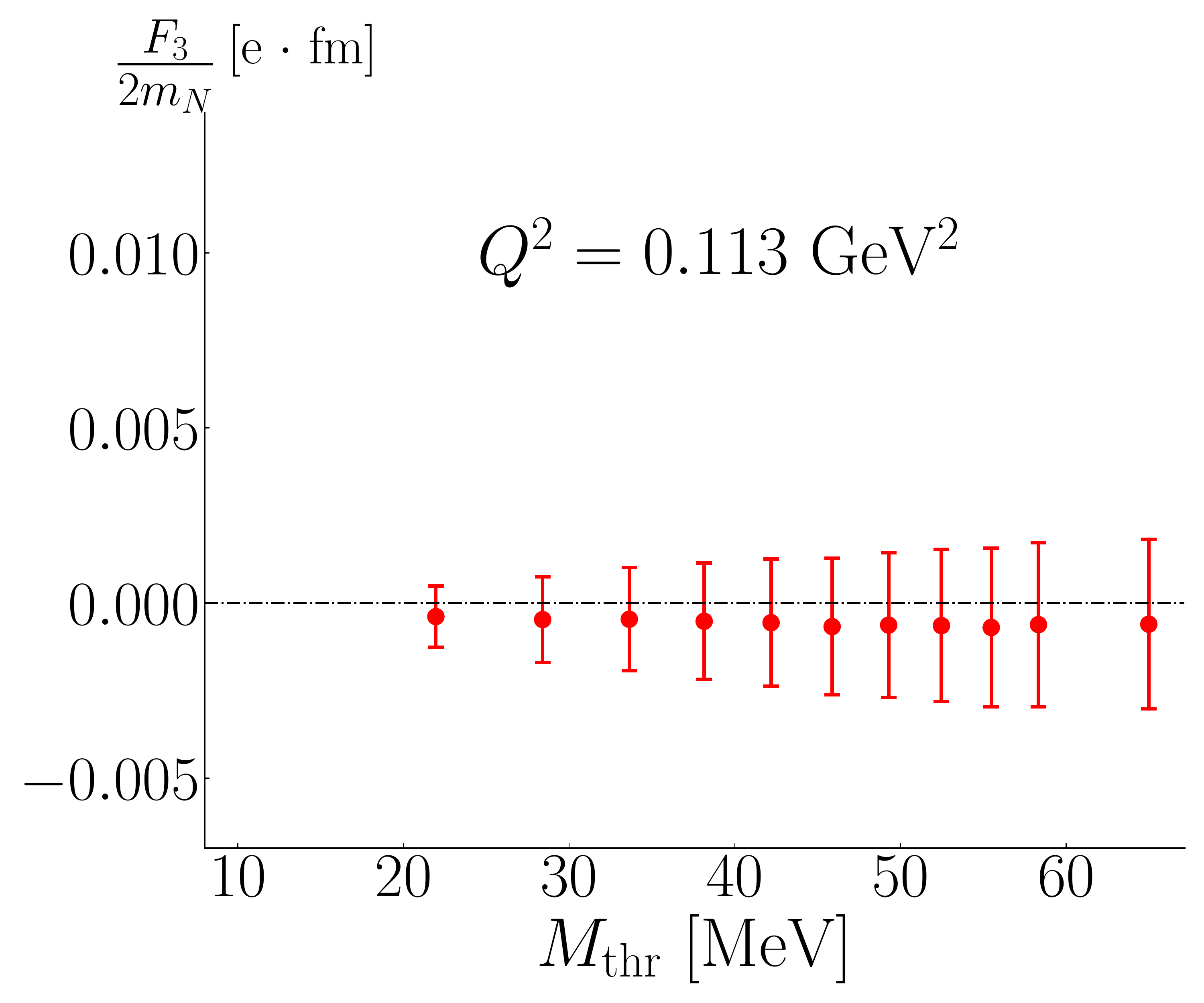}
\end{minipage}%
\hfill
\begin{minipage}[c]{0.33\linewidth}
\centering
\includegraphics[width=1.\textwidth]{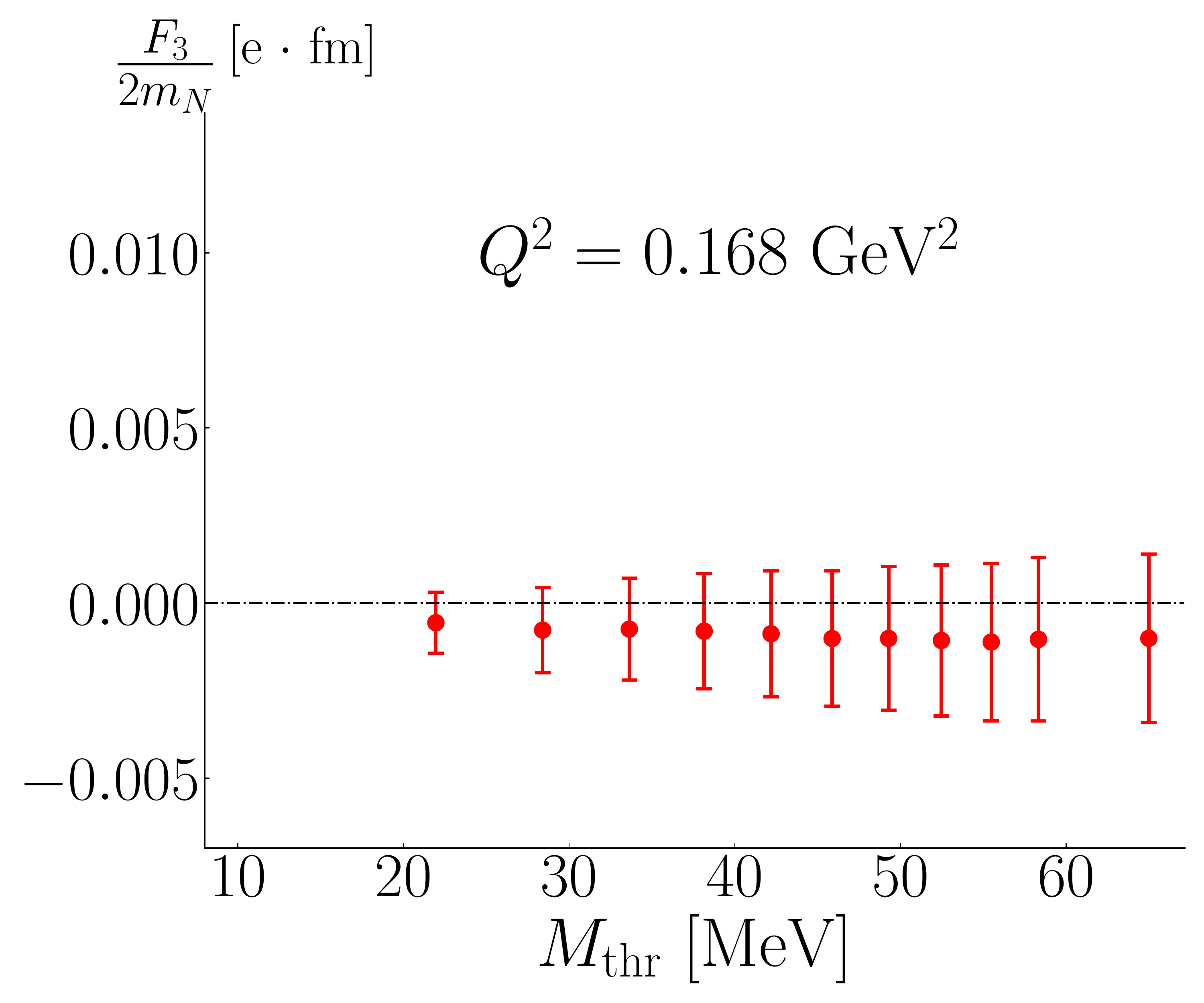}
\end{minipage}
\captionof{figure}{ Dependence of the \(F_3(0)\) on the smoothing scale $\tau_{\rm flow}$ for the gluonic (upper row) and cut-off \(M_{\rm thr}\) for the fermionic (bottom row) definitions used in computation of the topological charge, for the three smaller values of the momentum transfer squared.}
\label{fig:F3_d}
\end{figure}

As can be seen, the fermionic definition based on  spectral projectors of the topological charge provides a more accurate determination than the gluonic definition. Considering the absolute error as a bound for the magnitude of the nEDM, what we have is an improvement by a factor \(\sim 2\) when using the fermionic definition.  Therefore, the additional cost coming from the computation of the eigenmodes for the fermionic definition of \(\mathcal{Q}\), is compensated by the increased precision, leading to a large payoff in terms of the computational cost.
Moreover, the choice of maximum cut-off in the number of eigenmodes, i.e. \(M_{\rm thr}=64.98\) MeV, employed in the computation of the results presented, is in some sense the most conservative choice. Indeed, by looking at the dependence of the extrapolated value of \(F_3(0)\) as a function of \(M_{\rm thr}\), shown in the bottom row of Fig.~\ref{fig:F3_d}, it can be seen that  the mean value of the form factor does not depend on   \(M_{\rm thr}\) and only the error increases with increasing \(M_{\rm thr}\). This is in contrast with what is observed for the mixing angle $\alpha_N$, that instead changes significantly with \(M_{\rm thr}\). This is because the changes in  \(\alpha_N\) are small as compared to  the uncertainty of \(\Pi^{0k}_{3pt,\mathcal{Q}}\) and are thus not visible in \(F_3\) at the current level of precision.

    \begin{figure}[H]
    \centering
    \includegraphics[width=0.9\textwidth]{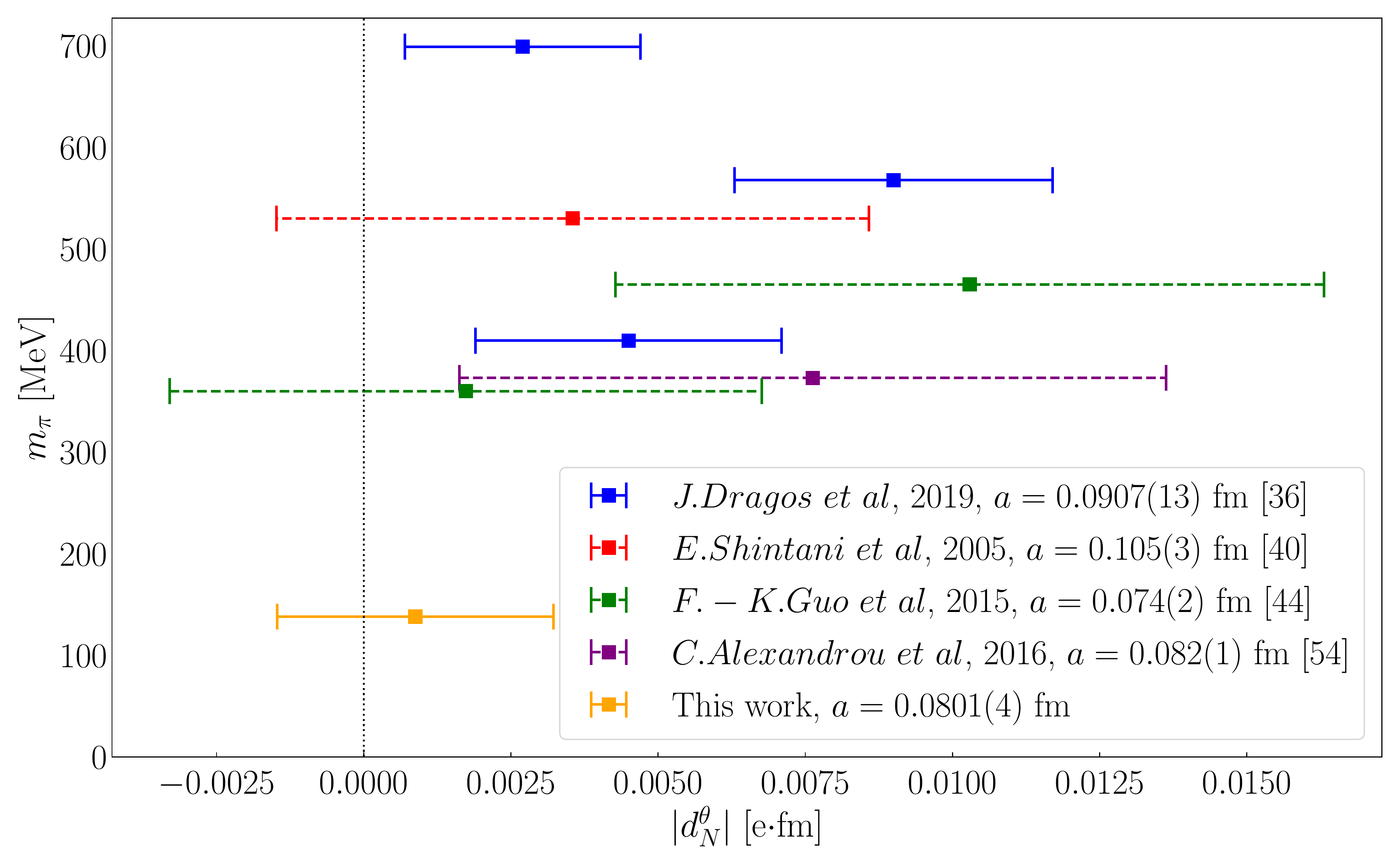}
    \captionof{figure}{Comparison with other lattice QCD determinations of nEDM present in literature. Values from Refs.~\cite{Shintani:2005xg,Guo:2015tla,Alexandrou:2015spa} (dashed error bars)  are not the ones from their original papers but are taken from Table~III of Ref.~\cite{Abramczyk:2017oxr}, where the spurious contribution coming from \(F_2(Q^2)\) is subtracted. See Ref.~\cite{Abramczyk:2017oxr} for further details.}
    \label{fig:comparison}
\end{figure}
\section{Comparison with other determinations of nEDM}
\label{sec:comparison}
In Fig.~\ref{fig:comparison} we provide a comparison of our result with those of other lattice QCD studies for a similar lattice spacing. We observe that our value has a statistical error that is comparable with the recent results in Ref.~\cite{Dragos:2019oxn} that were, however, computed at much larger values of the pion mass. Since the errors grow with decreasing pion mass and so does the computational cost, achieving such an accuracy it is a major outcome of this work. We note that the authors of Ref.~\cite{Dragos:2019oxn} using their results at these heavy pion masses perform a chiral extrapolation and find a non-zero value of \(\lvert d_N^\theta \rvert = 0.00152(71) \theta\:{\rm e}\cdot{\rm fm}\) at the physical point. The accuracy of their determination is due to the chiral extrapolation where systematic errors from using chiral expressions cannot be determined. It is worth mentioning that their actual data have uncertainties similar to this work. Our result computed directly at the physical point does not exclude a zero value, but provide a significant bound to the value of nEDM.

\section{Summary and Conclusions}
\label{sec:conclusions}

We compute the neutron electric dipole moment using an ensemble of $N_f =2+1+1$ twisted mass fermions simulated at the physical point and with a lattice spacing of $a \simeq 0.08$~fm. The extraction of the $CP$-violating form factor $F_3$ with the approach adopted in this work requires the calculation of the topological charge. For this reason we made use of a gluonic definition of the topological charge as well as a fermionic definition by means of gradient flow and spectral projectors, respectively. This enables us to test what effect different definitions of the topological charge may have on the nEDM.  $F_3(Q^2)$ cannot be extracted directly from the nucleon matrix element due to the presence of momentum appearing multiplicatively in front of the form factor.  The usual approach is to compute $F_3(Q^2)$ for finite $Q^2$ and then extrapolate to $Q^2 = 0$ using some fitting form. We would like to highlight the following three crucial aspects  of our investigation of the $CP$-odd form factor:

\begin{itemize}

    \item  {\it Determination of nEDM directly at the physical point}: We perform the computation directly at the physical value of the pion avoiding uncontrolled errors that a chiral extrapolation may result in.  We study $F_3(Q^2)$ as a function of the the momentum transfer $Q^2$, and show that there is no noticeable dependence on low values of $Q^2$ within our current statistical accuracy. We, thus, perform a weighted average on the three lowest ones to extract the value of the form factors at $Q^2=0$.

    \item  {\it Definition of the topological charge}:  The approach is based on using spectral projectors to determine  the topological susceptibility yielding a statistical error that is  twice smaller than when using a  gluonic definition whether using gradient flow, link smearing, or cooling are applied~\cite{Alexandrou:2017hqw, Alexandrou:2017bzk}.  Furthermore, it has been shown that the topological susceptibility calculated using the  spectral projectors approach shows milder  lattice artifacts than  when using a gluonic definition \cite{Alexandrou:2017bzk}. Both these observations have led us  to investigate the approach based on spectral projectors for the definition of the topological charge entering  the evaluation of the nEDM since we expected that these features would also  hold  by secondary quantities involving the topological charge. The necessity to reduce as much as possible the lattice cut-off effects stem from the fact that currently we do not have three ensembles with different lattice spacings  at the physical point  to enable us to take the continuum limit. 
   Therefore, while $F_3(0)$ extracted using the fermionic definition is  consistent with the value extracted  using the gluonic definition, the statistical errors of the former are about half as compared to those when using the gluonic definition and the cut-off effects are expected from the above consideration to be  milder. We, thus, quote as our final value of nEDM the one resulting from using spectral projectors to define the topological charge. We find
    
    \begin{eqnarray}
    \ \lvert d_N^{\theta}\rvert = 0.0009(24) \; \theta \; e , {\rm fm}
    \end{eqnarray}
    which is compatible with zero.
    
    There is a subtlety that we would like to clarify at this point, namely the fact that we do not observe a clear plateau for $\alpha_N$ when using spectral projectors as a function of $M_{\rm thr}$.  This is an artifact of carrying out the investigation at one lattice spacing. Only after a continuum extrapolation keeping the  renormalised cut-off $M_{\rm thr}$ in physical units fixed one can extract physical observables  that are independent of the cut-off. The same holds for the field theoretic definition for which the gradient flow time in physical units must be fixed as one takes the continuum limit. This is illustrated in Fig.~\ref{fig:susc_fg} for the susceptibility, where only after the continuum limit is taken there is an agreement among the approaches.
    
    A question arises as of why when using the gradient flow with the Wilson smoothing action we obtain quantities such as the topological susceptibility and $\alpha_N$ that have a very mild dependence on the scale $\tau_{\rm flow}$ leading to a fast convergence to a plateau. This can be understood within the following context. On the semi-classical level, gradient flow smooths out small instantons corresponding to ultra-violet (UV) contributions. This happens even at small  flow times and, thus, we are left with a topological content which remains unaltered until the smoother starts affecting the large instantons contributing to the topological structure of the theory. On the other hand, the same quantities extracted using the spectral projectors reveal a behaviour that depends much stronger on the associated cut-off. This is because, for a theory that breaks chiral symmetry, as we sum up eigenvalues of the squared Dirac Matrix, the summation will keep increasing. This observation is fully supported by the results on the topological susceptibility demonstrated in Ref.~\cite{Alexandrou:2017bzk}. One may argue that the existence of a plateau enables to reliably quote a value of a quantity at a given lattice spacing; this is of course true but has no physical importance since in the end a continuum extrapolation is needed in order to get rid of  discretization effects. Nevertheless, we observe that, when using spectral projectors, $F_3$ appears to exhibit a plateau versus $M_{\rm thr}$ coupled with the extra benefit of reduction in the statistical uncertainty.

    \item {\it Alternative approaches of extracting the $\theta$-induced nEDM.} The current investigation reveals the difficulty of this particular method to deliver a statistically significant result.  Practically, one needs to increase measurements of at least an order of magnitude to reduce the error significantly hopping to exclude a zero value.   Employment of techniques, such as volume clustering~\cite{Liu:2017man,Syritsyn:2019vvt} or spectral projectors as a definition of the topological charge as done in this work, help but do not solve the problem. 
    
    Another possibility is the investigation of the nEDM using configurations generated with an imaginary $\theta$-term. Since the nEDM depends on contributions from non-trivial topological sectors, introducing a dynamical $\theta$-term one may improve the importance sampling for the nEDM signal inducing nonzero average topological charge. Exploratory investigations using this method with a field theoretic definition of the topological charge are under way.
\end{itemize}

\section{Acknowledgements}
We would like to thank all the members of the Extended Twisted Mass Collaboration for a most conducive cooperation.  In particular, we express our gratitude  to K. Cichy for providing the data on the susceptibility. We thank M. Constantinou for providing us the value of $Z_S$.  We also  thank M.~D'~Elia and C.~Bonanno for valuable discussions.
A.A. is financially supported by the European Union's Horizon 2020 research and innovation programme ``Tips in SCQFT'' under the Marie Sk\lpol odowska-Curie grant agreement No. 791122.   K.H. is financially supported by the Cyprus Research  and Innovation  foundation  under  contract  number  POST-DOC/0718/0100. A.T. is a Marie Sklodowska-Curie fellow funded by the European Union’s Horizon 2020 research and innovation programme under grant agreement No 765048. Results were obtained using Piz Daint at Centro Svizzero di Calcolo Scientifico (CSCS), via the project with id s702. We acknowledge PRACE for awarding us access to Marconi100 at CINECA, Italy,  where part of our work is carried out within the project with Id Pra20\_5171. This work also used computational resources from the John von Neumann-Institute for Computing on the Juwels system at the research center in J\"{u}lich, under the project with id CHCH02.

\printbibliography[heading=bibintoc]

\end{document}